\documentclass[aps,pre,twocolumn,showpacs,superscriptaddress]{revtex4-1}  
\usepackage{graphicx}  
\usepackage{dcolumn}   
\usepackage{bm}        
\usepackage{amssymb}   

\usepackage{txfonts}
\usepackage{hyperref}
\usepackage{graphicx}
\usepackage{subfigure}
\usepackage{pdfpages}

\begin{document}

\title{Simultaneous first and second order percolation transitions in
interdependent networks}

\author{Dong Zhou}
\affiliation{Department of Physics, Bar-Ilan University, Ramat Gan 52900, Israel}
\author{Amir Bashan}
\affiliation{Department of Physics, Bar-Ilan University, Ramat Gan 52900, Israel}
\author{Reuven Cohen}
\affiliation{Department of Mathematics, Bar-Ilan University, Ramat Gan 52900, Israel}
\author{Yehiel Berezin}
\affiliation{Department of Physics, Bar-Ilan University, Ramat Gan 52900, Israel}
\author{Nadav Shnerb}
\affiliation{Department of Physics, Bar-Ilan University, Ramat Gan 52900, Israel}
\author{Shlomo Havlin}
\affiliation{Department of Physics, Bar-Ilan University, Ramat Gan 52900, Israel}


\date{\today}

\begin{abstract}
In a system of interdependent networks, an initial failure of nodes invokes a cascade
of iterative failures that may lead to a total collapse of the whole system in a form
of an abrupt first order transition. When the fraction of initial failed nodes $1-p$
reaches criticality, $p=p_c$, {the abrupt collapse occurs by spontaneous
cascading failures. At this stage,} the giant component decreases slowly in a plateau form
and the number of iterations in the cascade, $\tau$, diverges. {The origin
of this plateau and its increasing with the size of the system {remained}
unclear. Here we find that simultaneously with the abrupt {first order transition}
a {spontaneous} second order percolation occurs {during the cascade
of iterative failures}. This sheds light on the origin of the plateau and
{on how its} length scales with the size of the system.} Understanding the
{critical nature of the} dynamical process of cascading failures
{may be useful} for designing strategies for preventing and
mitigating catastrophic collapses.
\end{abstract}

\maketitle

\section{Introduction}

Interdependent network systems attract a growing interest in the last years
\cite{LAP07,ROS08,PAN08,VES10,BUL10A,PAR10A,GAO11,LEI09,MOR12,SAU12,SHA11,GAO10,HUA11,GOM13,AGU13,CHA12,BIA13,CEL13,RAD13,LI12,BAS13}.
They represent real world systems
composed of different types of interrelations, connectivity links between entities
(nodes) of the same network to share supply or information and dependency links
which represent a dependency of one node on the function of another node in
another network. Consequently, failure of nodes may lead to two different effects:
removal of other nodes from the same network which become disconnected from the
giant component and failure of dependent nodes in other networks. The synergy
between these two effects leads to an iterative chain
cascading of failures. Buldyrev \textit{et al} \cite{BUL10A} show that, in a system
of two fully interdependent random networks, when the fraction of failed nodes $1-p$ is
smaller than a critical value, $p>p_c$, the cascading failures stop after some
iterations and a finite fraction of the system, $P_\infty>0$, remains functioning
and connected to the giant component. A larger initial damage, $p<p_c$, invokes a
cascading failure that fragments the entire system and $P_\infty=0$. Thus, when $p$
approaches $p_c$ from above, the giant component, $P_\infty$, discontinuously jumps
to zero in a form of a first order transition. The number of iterations in the
cascade, $\tau$, diverges when $p$ approaches $p_c$, a behavior that was suggested as
a clear indication for the transition point in numerical simulations \cite{PAR11}.

\begin{figure*}
 \includegraphics[width=0.8\textwidth]{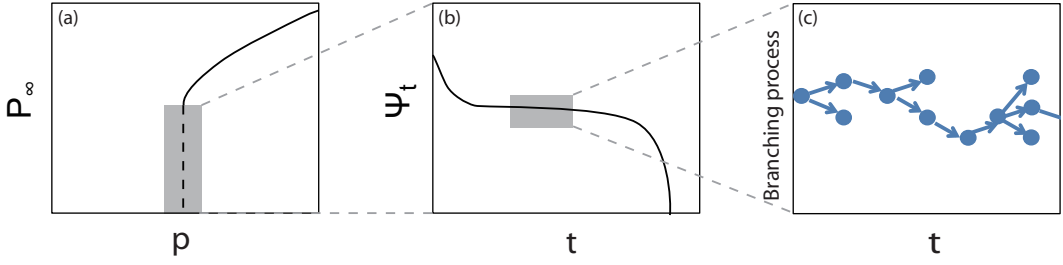}
 \caption{{{Demonstration of} the simultaneous first and second order
 transitions in cascading failures of interdependent networks. At the critical point $p_c$,
 (a) the mutual giant component has a sudden jump to zero, while (b) the dynamical process
 of cascading failures is governed by a long plateau stage. In this plateau stage, a second
 order percolation occurs, which is (c) characterized by a random branching process at criticality,
 i.e., average branching factor is one (see Fig. \ref{fig3:subfig:b}).}}
 \label{fig1}
\end{figure*}

{Among the main features found are the collapse of the system with time
{(steps of cascading failures)} in a plateau form (see Fig. {\ref{fig1}}), and the
increase of the plateau length with the system size. Although this phenomena was
observed in different models and in real data, its origin {remained} unclear \cite{BUL10A}.
To understand the origin of this phenomena we focus on fully interdependent
Erd\H{o}s-R\'{e}nyi (ER) networks. Surprisingly, we find here that during the abrupt
collapse there appears a hidden spontaneous second order percolation transition that
controls the cascading failures, as demonstrated in Fig. \ref{fig1}. We show here
that this simultaneous second order phase transition{, characterized by
long branching trees near criticality,} is the origin of the observed long plateau regime in the cascading
failures and its dependence on system size. Moreover, the second order
transition sheds light on the critical behavior observed in the collapse of real
world systems such as the power law distribution of blackout sizes \cite{CAR04A,HOL06,BAK06,ANC05}.}

We {also find, as a result of this new understanding,} that even though the {mean-field
(MF)} approximations are found to be accurate in predicting $p_c$ and $P_\infty$,
it {does not} represent the dynamical process of cascading failures near criticality.
{This is since,} the critical dynamics is strongly affected by random
fluctuations {due to the second order transition} which are
not considered in the MF approach. We study the effect of these fluctuations
on the total number of iterations $\tau$ at criticality and find that its average and standard
deviation scale as $N^{1/3}$, in contrast to the MF prediction of
$\langle\tau\rangle\sim N^{1/4}$ \cite{BUL10A}. We present a theory for the dynamics
at criticality, which explains the origin of this difference.

\section{Model of Interdependent Networks}

In the fully interdependent networks model, $A$ and $B$ are two networks of
the same size $N$. Each $A$-node $a_{i}$ depends on exactly one randomly-chosen
$B$-node $b_{j}$, and $b_{j}$ also only depends on $a_{i}$. The initial
attack is removing randomly a fraction $1-p$ of $A$-nodes in one network. Nodes in one network
that depend on removed nodes in the other network are also removed, causing a
cascade of failures. As nodes and edges are removed, each network breaks up into
connected components (clusters). {It is known that for single random networks, there is
at most one component (giant component) which occupies a finite
fraction of all $N$ nodes (see \cite{COHbook}).} We assume that only nodes belonging
to the giant component connecting a finite fraction of the network are still
functional. Since the two networks have different topological structures, the
failure will spread as a cascading process in the system \cite{PAR10B,HU13,CEL13}.
{Here, one time step means that dependency failures and percolation failures occur at a given iteration in
networks $A$ and $B$ respectively, and each network reaches a new smaller giant component.}

The MF theory of this model with ER networks with average degrees
$k_{A}$ and $k_{B}$ has been developed using generating functions of the
degree distribution. This theory predicts the giant
component size as a function of $p$, and accurately evaluate the
first order phase transition threshold $p^{MF}_{c}$ for the infinite-size
system. In fact, each realization in the simulation has its own critical
threshold which we denote by $p_{c}$. {In this paper, a new realization means that we
generate networks $A$ and $B$ again, as well as the interdependency
links, and then we perform the initial attack according to a new random
attack order (see \footnote{We tested also by taking a few different realizations
and keeping the network structures but changing only the attack order.
The results have been found to be very similar, as seen in Fig. \ref{fig8}}).} Note that for $N\to\infty$,
$p_c$ values in single realizations are the same and equal to $p_c^{MF}$.

\section{Scaling Behavior in the Critical Dynamics}

Here, we investigate the dynamics of the critical cascading failures for each
single realization {of a pair of finite coupled networks. For simplicity, networks $A$
and $B$ have the same average degree $k$.} The value of $p_{c}$ of each realization,
can be determined accurately by randomly removing nodes one by one {until} the
system fully collapses.

\begin{figure*}
 \centering
 \subfigure{
  \label{fig2:subfig:a}
  \includegraphics[width=0.295\textwidth]{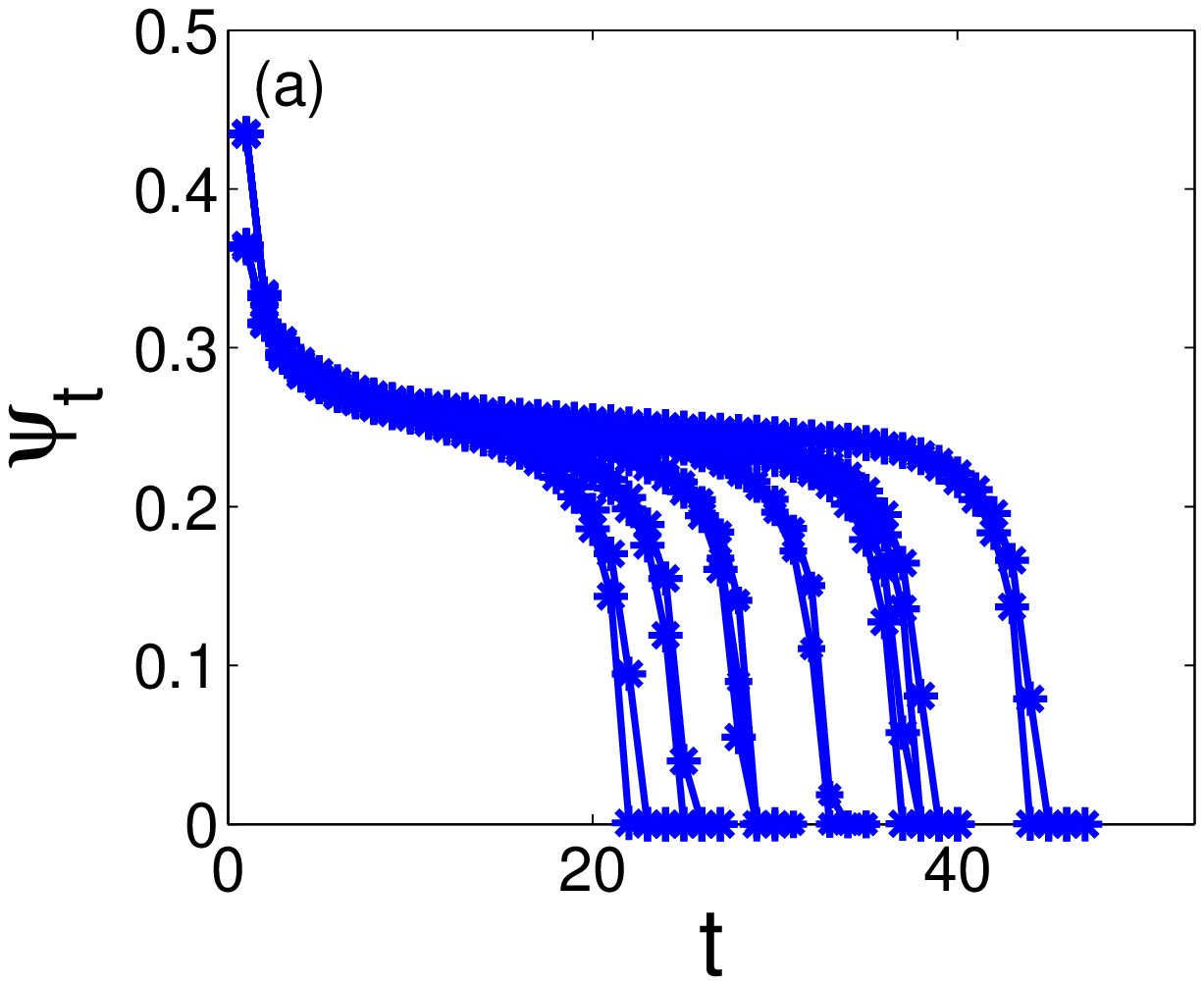}}
 \subfigure{
  \label{fig2:subfig:b}
  \includegraphics[width=0.295\textwidth]{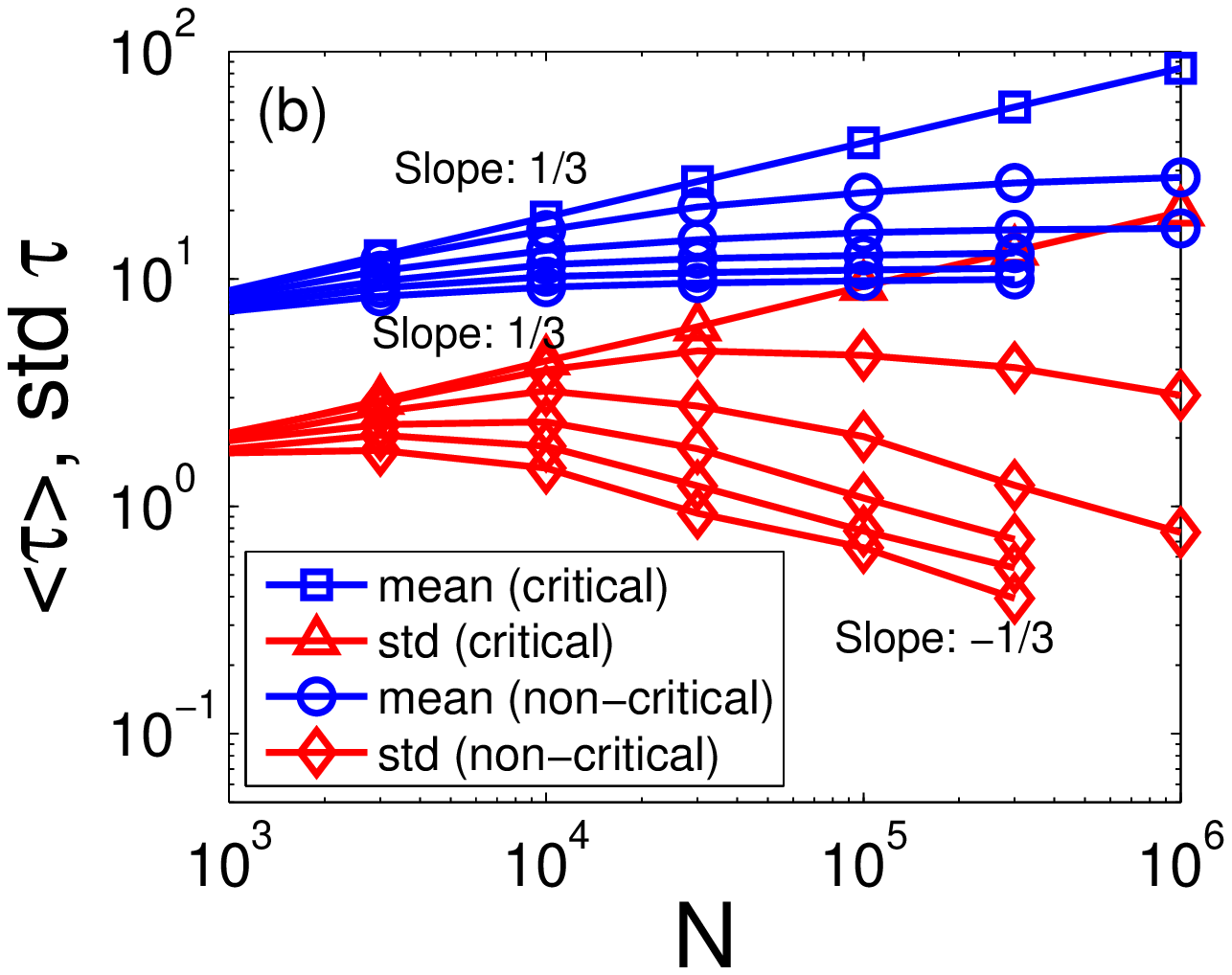}}
  \subfigure{
  \label{fig2:subfig:c}
  \includegraphics[width=0.295\textwidth]{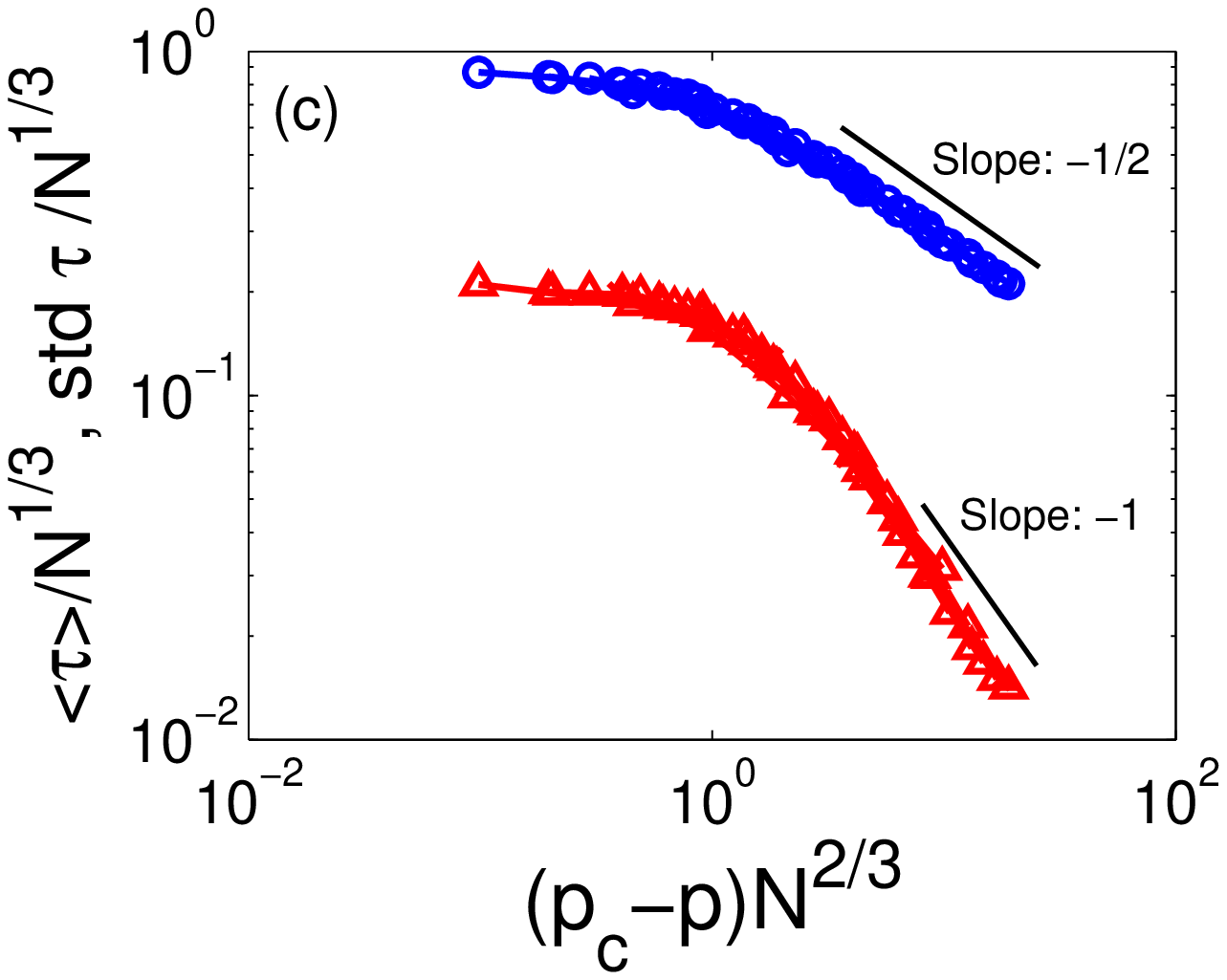}}
 \caption{(Color online) \textbf{a}. Dynamical process of the giant component size $\psi_{t}$ of
 network $A$ in simulation at $p_{c}^{MF}$ (15 realizations).
 \textbf{b}. Scaling behavior of the mean (blue) and the standard deviation (red) of the total
 time $\tau$ at $p_{c}$ (critical) or below $p_{c}$ (non-critical) for each realization. Each curve
 here corresponds to a fixed value of $p$. \textbf{c}. The scaled version of (b).
 We consider for Fig. \ref{fig2:subfig:a} the case $N=300\,000$ and $k=5$ with
 $15$ realizations. For Figs. \ref{fig2:subfig:b} and \ref{fig2:subfig:c}, we have
 $k=5$ for the different $N$ values we analyzed. {Each point here is
 the mean or the standard deviation over} $200$ realizations of $N=10^6$, and order
 of $10^4$ realizations for $N\leq 300\,000$.}
 \label{fig2}
\end{figure*}

Fig. \ref{fig2:subfig:a} exhibits several realizations of simulations at
{$p_{c}^{MF}$}. As seen
at criticality, the total time $\tau$ has large
fluctuations. Each realization has a stage of time steps
(a plateau) where the giant component of network $A$ decreases very slowly.
Before or after this plateau stage, the cascading failure process is much
faster.

Fig. \ref{fig2:subfig:b} and Fig. \ref{fig2:subfig:c} show the scaling
behaviors of the mean and the standard deviation of $\tau$ as a function
of $N$ and $p-p_{c}$. In our simulations, we consider $p\leq p_{c}$, and
only those realizations that fully collapse. We wish to understand how $N$
and $p-p_{c}$ affect the mean and the standard deviation of the total time
$\tau$.

It can be seen from Fig. \ref{fig2:subfig:b} that $\langle\tau\rangle$ increases
with $N$ as $\langle\tau\rangle\sim N^{1/3}$ at $p=p_{c}$. However, when $p<p_{c}$,
$\langle\tau\rangle$ becomes constant for large values of $N$. Thus, we assume
the following scaling function,

\begin{equation}
\label{eq1}
 \langle\tau\rangle\sim N^{1/3}\cdot f(u),
\end{equation}

\noindent where $u=(p_{c}-p)\cdot N^{1/\alpha}$, and $f(u)$ is a function
which satisfies: $f(u)\sim const.$ for $u<<1$, $f(u)\sim u^{-\alpha/3}$
for $u>>1$, and we determine $\alpha$ such that the best scaling occurs.

To test Eq. (\ref{eq1}) and identify $\alpha$, we plot in Fig. \ref{fig2:subfig:c} $\langle\tau\rangle/N^{1/3}$
versus $u$. We find that the best choice of $\alpha$ for obtaining
a good scaling collapse is $\alpha=3/2$. In this way, we can see that the slope
of each curve changes from $0$ to about $-1/2$ at $u=(p_{c}-p)\cdot N^{2/3}\approx 1$.
Therefore, the scaling behavior of $\langle\tau\rangle$ for $N<N^{\ast}\sim(p_{c}-p)^{-3/2}$ is

\begin{equation}
 \label{eq2}
 \langle\tau\rangle\sim N^{1/3},
\end{equation}

\noindent independent of $p$ (Fig. \ref{fig2:subfig:b}). This means that system
sizes of $N<N^{\ast}$ are at criticality even though $p<p_{c}$. For $N>N^{\ast}$,
{$\langle\tau\rangle\sim N^{1/3}\cdot u^{-1/2}=(p_{c}-p)^{-1/2}$,} independent of
$N$ (Fig. \ref{fig2:subfig:b}) (non-critical behaviors).
This yields the crossover {$N^{\ast}\sim(p_{c}-p)^{-\alpha}=(p_{c}-p)^{-3/2}$,}
between the critical behavior for $N<N^{\ast}$ and non-critical
for $N>N^{\ast}$. For $p\rightarrow p_{c}$, $N^{\ast}\rightarrow\infty$
and for all $N$ one observes the critical behavior. The crossover system size,
$N^{\ast}$, can be regarded as a correlation size analogously to the correlation
length in regular percolation \cite{BUNbook,STAbook}.

Fig. \ref{fig2:subfig:b} also illustrates the scaling behaviors of the
standard deviation, $\textup{\textbf{std}}(\tau)$. For $p=p_{c}$, we obtain
$\textup{\textbf{std}}(\tau)\sim N^{1/3}$, i.e., it increases as the
same rate as the mean. However, for $p<p_{c}$, the slope in the right tail of $\textup{\textbf{std}}(\tau)$
in Fig. \ref{fig2:subfig:b} is about $-1/3$. Thus, we assume a scaling function
for $\textup{\textbf{std}}(\tau)$:
\begin{equation}
 \label{eq5}
 \textup{\textbf{std}}(\tau)\sim N^{1/3}\cdot g(u),
\end{equation}
where $u=(p_{c}-p)\cdot N^{1/\alpha}$, and $g(u)$ satisfies: $g(u)\sim const.$ for $u<<1$,
and $g(u)\sim u^{-2\alpha/3}=u^{-1}$ for $u>>1$.

Fig. \ref{fig2:subfig:c} shows that the scaling behavior of $\textup{\textbf{std}}(\tau)$
assumed in Eq. (\ref{eq5}) is supported by simulations with the best choice $\alpha=3/2$
as for $\langle\tau\rangle$. The slope of the right tail in Fig. \ref{fig2:subfig:c} is
indeed $-1$. Thus, for $N<N^{\ast}$, we have the critical behavior:
{$\textup{\textbf{std}}(\tau)\sim N^{1/3}$;} and for $N>N^{\ast}$,
{$\textup{\textbf{std}}(\tau)\sim N^{1/3}\cdot u^{-1}=N^{-1/3}(p_{c}-p)^{-1}$.}
Thus, we have the non-critical behavior also consistent with Fig. \ref{fig2:subfig:b}.

\section{The Spontaneous Second Order Percolation Transition}

Next we explore the mechanism behind the scaling behaviors near $p_{c}${.
We show that it is due to a spontaneous second order percolation transition} and
explain the deviation from the MF theory. The failure size, $s_t$, the number of $A$-nodes that
fail at time step $t$, {during the plateau from the coupled networks system,} is a zero fraction of the network size $N$. This is supported
by simulations shown in Fig. \ref{fig4:subfig:a}. We regard each node that fails
due to dependency at the beginning of the
plateau stage as a root, $a_{i}$, of a failure tree {(see Fig. \ref{fig1})}. After that, the removal of each
root $a_{i}$ will cause the failure of several other $A$-nodes due to percolation.
Then, several $B$-nodes will fail due to dependency and percolation in network
$B$. At the next time step, several $A$-nodes further fail due to dependency
and percolation, which can be regarded as the result of the original removal of
the root node $a_{i}$. Notice that the failures in network $A$ caused
by removing different single nodes $a_{i}$ have very few overlaps due to the
randomness and the large size of $N$. Therefore, we
can describe the plateau stage by the growth of all these independent failure trees
with the branching factor $\eta_{t}=s_{t+1}/s_{t}$.

\begin{figure}
 \subfigure{
  \label{fig3:subfig:a}
  \includegraphics[width=0.232\textwidth]{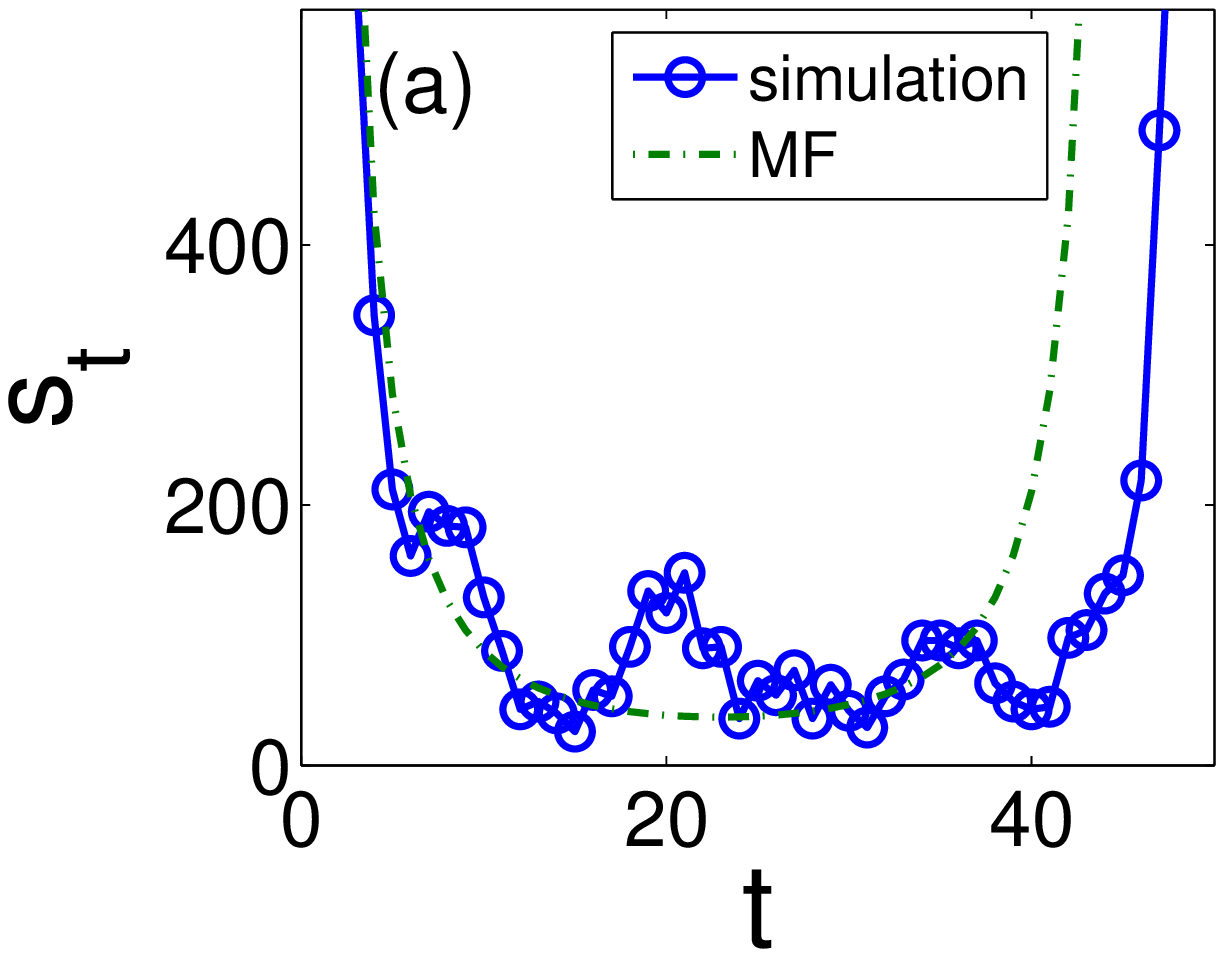}}
 \subfigure{
  \label{fig3:subfig:b}
  \includegraphics[width=0.232\textwidth]{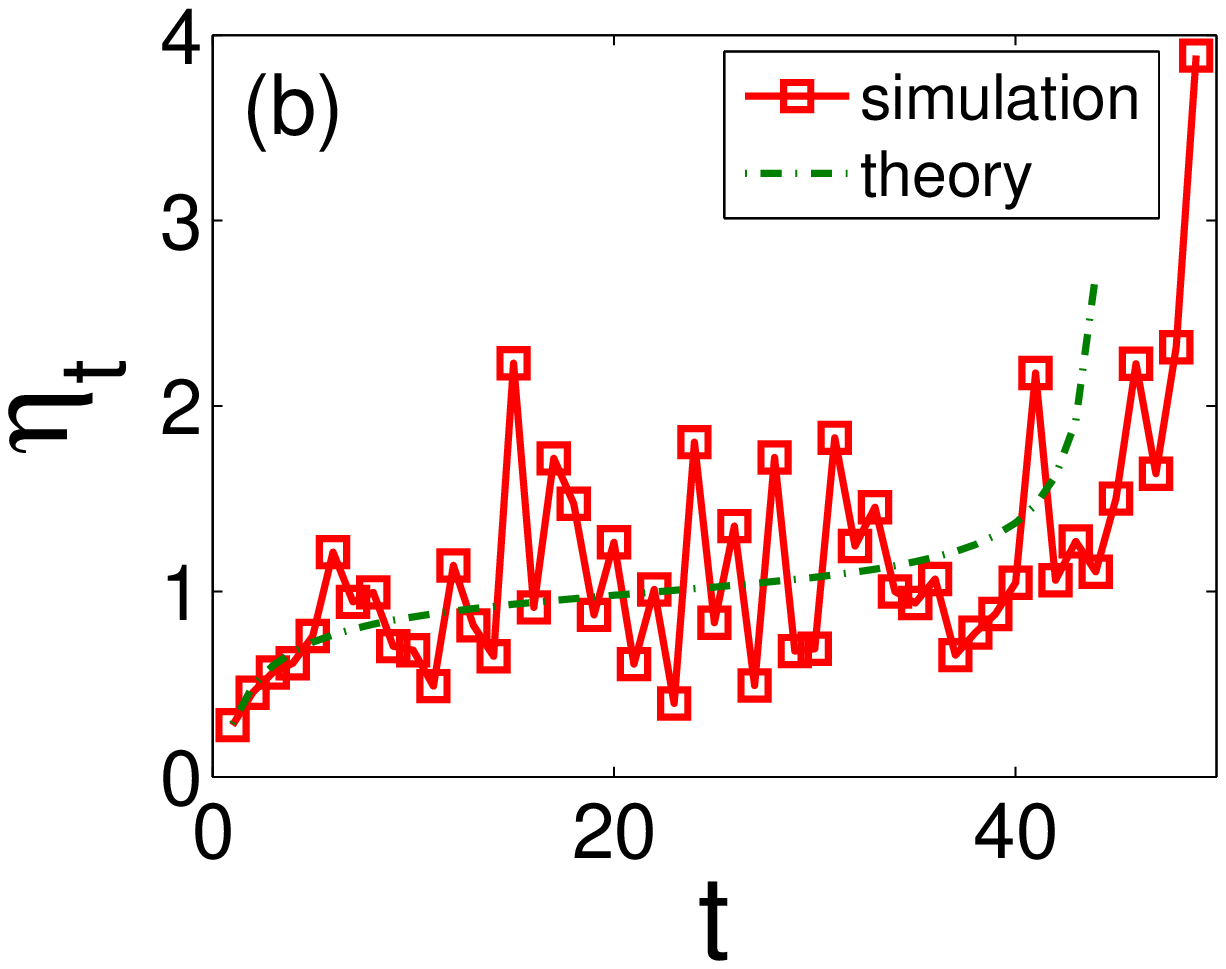}}
   \subfigure{
  \label{fig3:subfig:c}
  \includegraphics[width=0.232\textwidth]{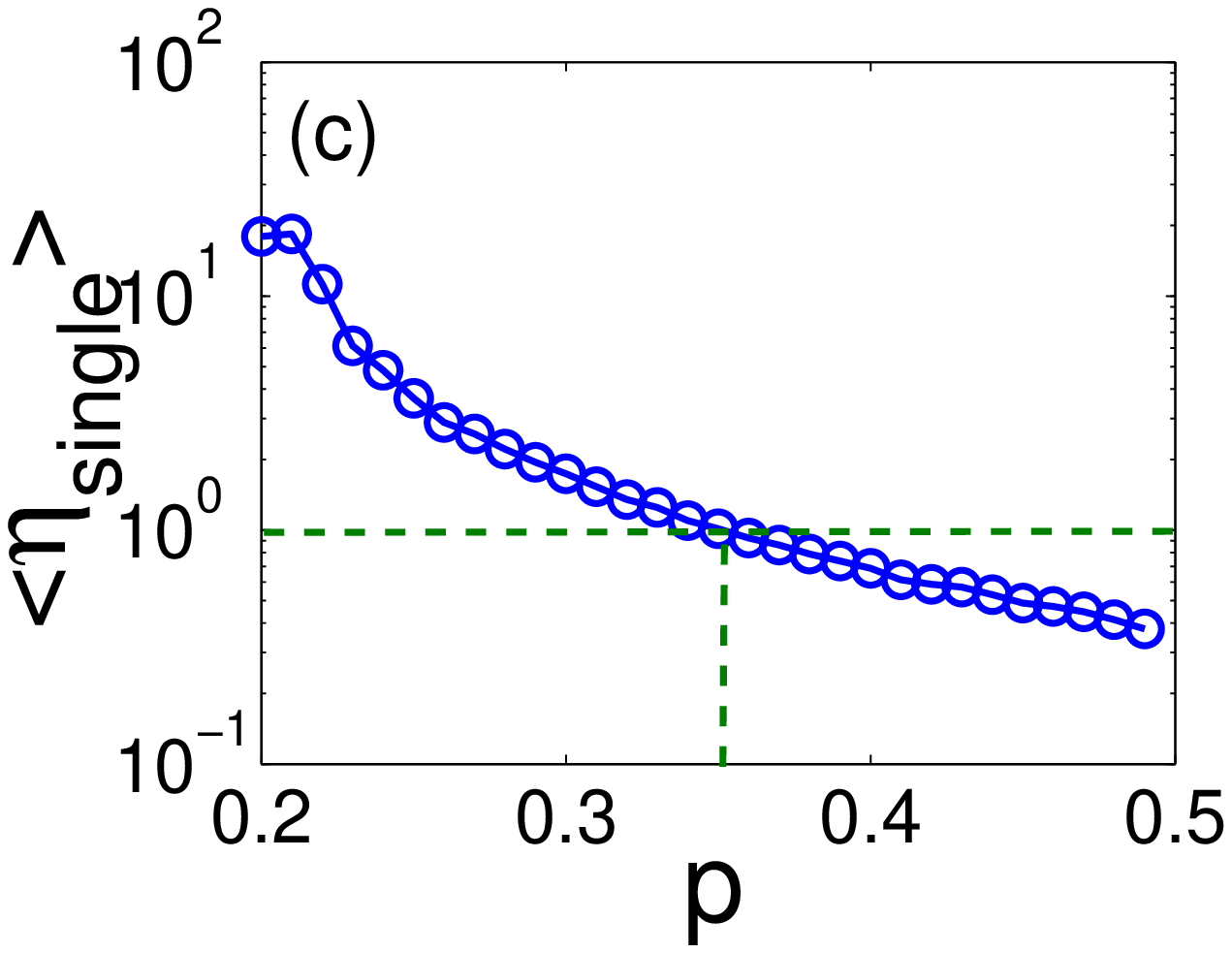}}
  \subfigure{
  \label{fig3:subfig:d}
  \includegraphics[width=0.232\textwidth]{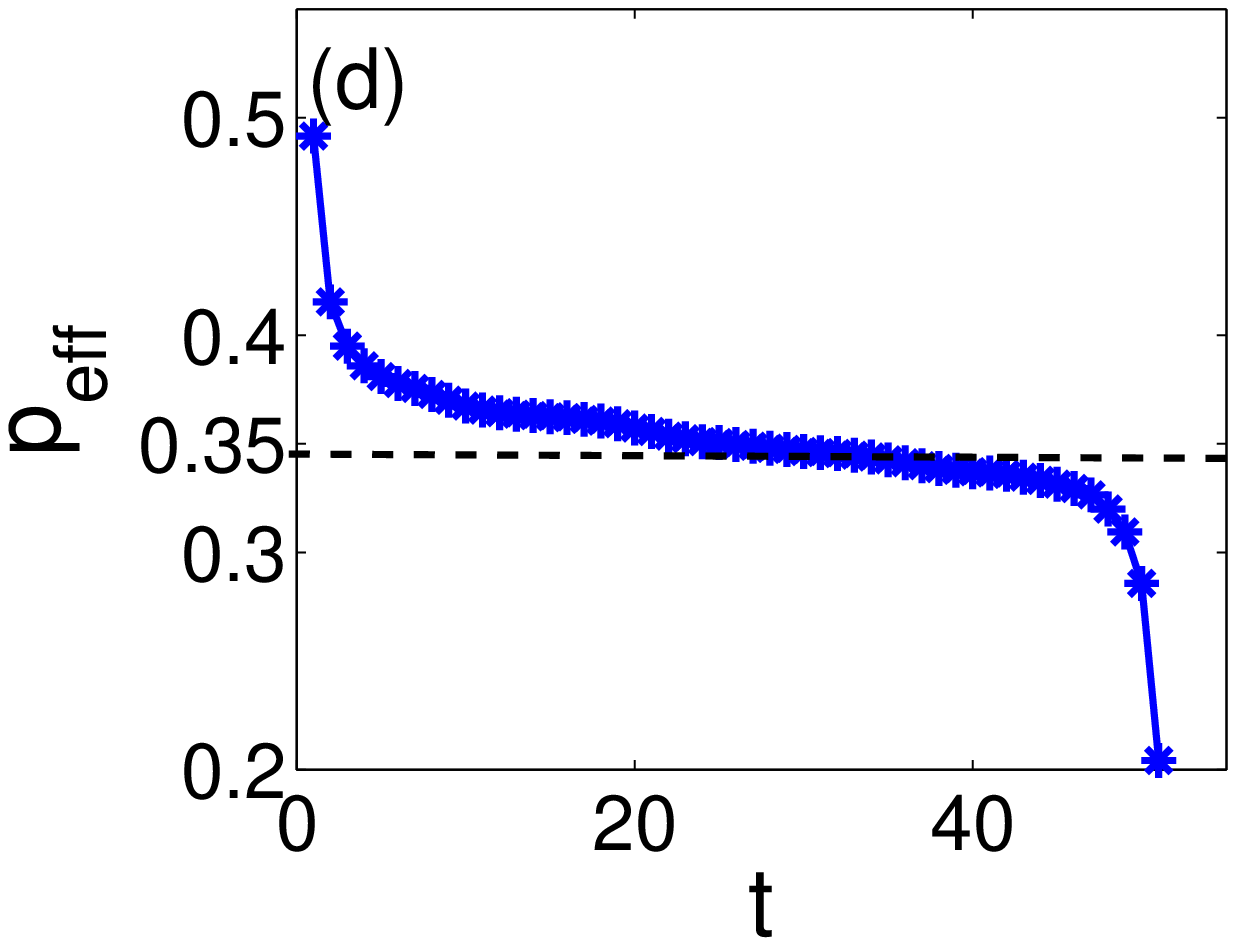}}
 \caption{(Color online) \textbf{a}. The (blue) line with circles is the variation of failure sizes $s_{t}$
 (only the plateau stage) for one realization in simulation. Here, $k=5$, $p=p_{c}^{MF}$ and $N=50\,000$.
 The (green) dashed dotted line shows $s_{t}$ for the MF case for $k=5$, $N=50\,000$, and
 $p=2.454/k$ slightly below $p^{MF}_{c}=2.4554/k$. \textbf{b}. The
 (red) line with rectangles shows the variation of the average branching factor $\eta_{t}$
 for one realization in simulation. Here, $k=5$, $p=p_{c}^{MF}$ and $N=50\,000$. The (green)
 dashed dotted line shows $\eta_{t}$ of the analytic MF solution. Here, $k=5$, $N=50\,000$ and
 $p=2.4536/k$ below $p^{MF}_{c}$. {In both (a) and (b), the MF values have similar behaviors as the
 simulation values, but the MF curves are smooth and show no fluctuations.} \textbf{c}. The average
 branching factor $\langle\eta_{single}\rangle$ for different values of $p$ on a single
 ER network. Here, $k=5$, $N=250\,000$ for $3000$ realizations. A threshold $\tilde{p}$
 where $\langle\eta_{single}\rangle=1$ can be observed at $p=\tilde{p}\approx 0.35$. \textbf{d}. The variation of the
 effective $p$ for one realization in the simulations. Here, $k=5$, $p=p_{c}^{MF}=2.4554/k$ and $N=50\,000$. }
 \label{fig3}
\end{figure}

\begin{figure}
 \centering
 \subfigure{
  \label{fig4:subfig:a}
  \includegraphics[width=0.23\textwidth]{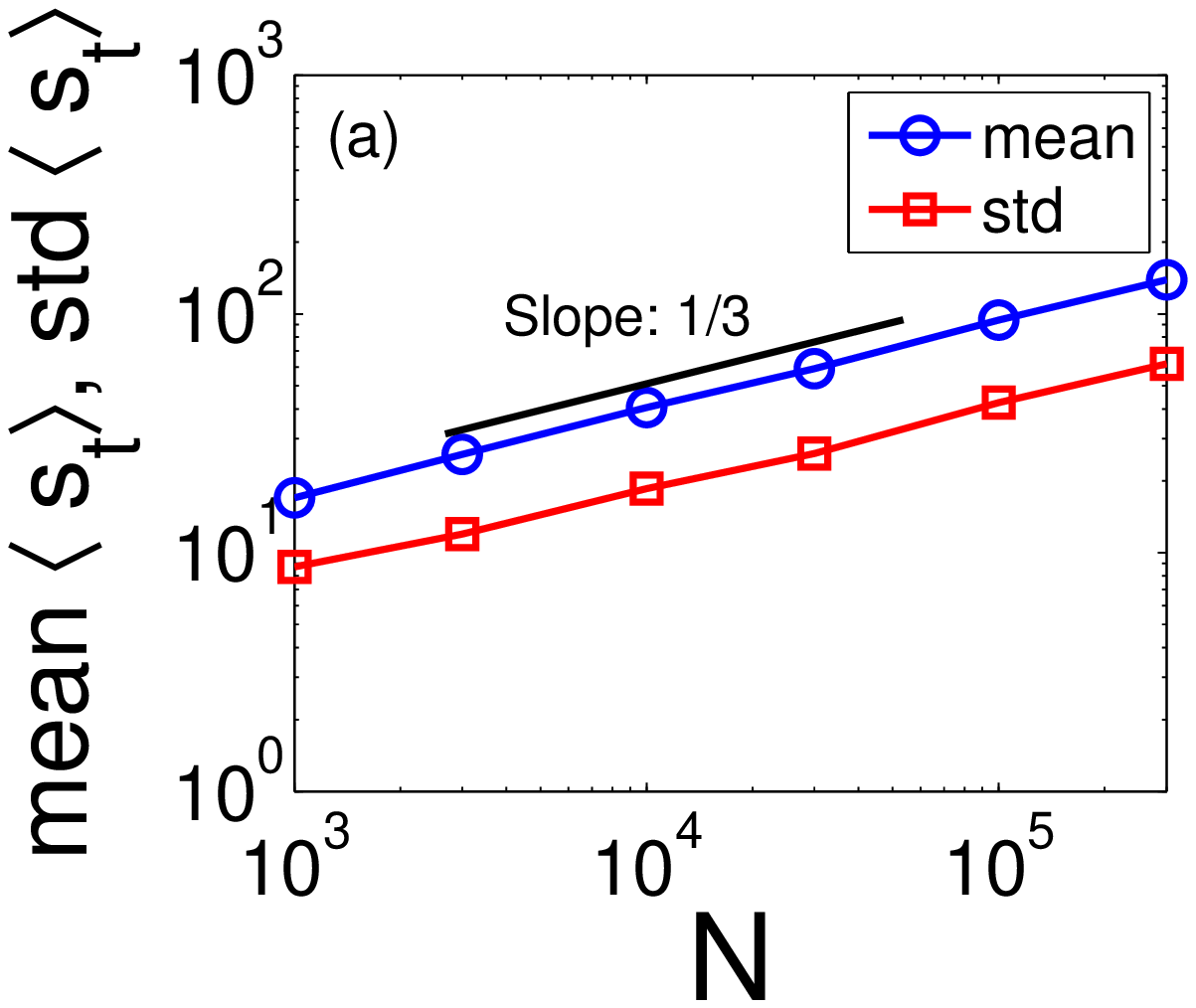}}
 \subfigure{
  \label{fig4:subfig:b}
  \includegraphics[width=0.23\textwidth]{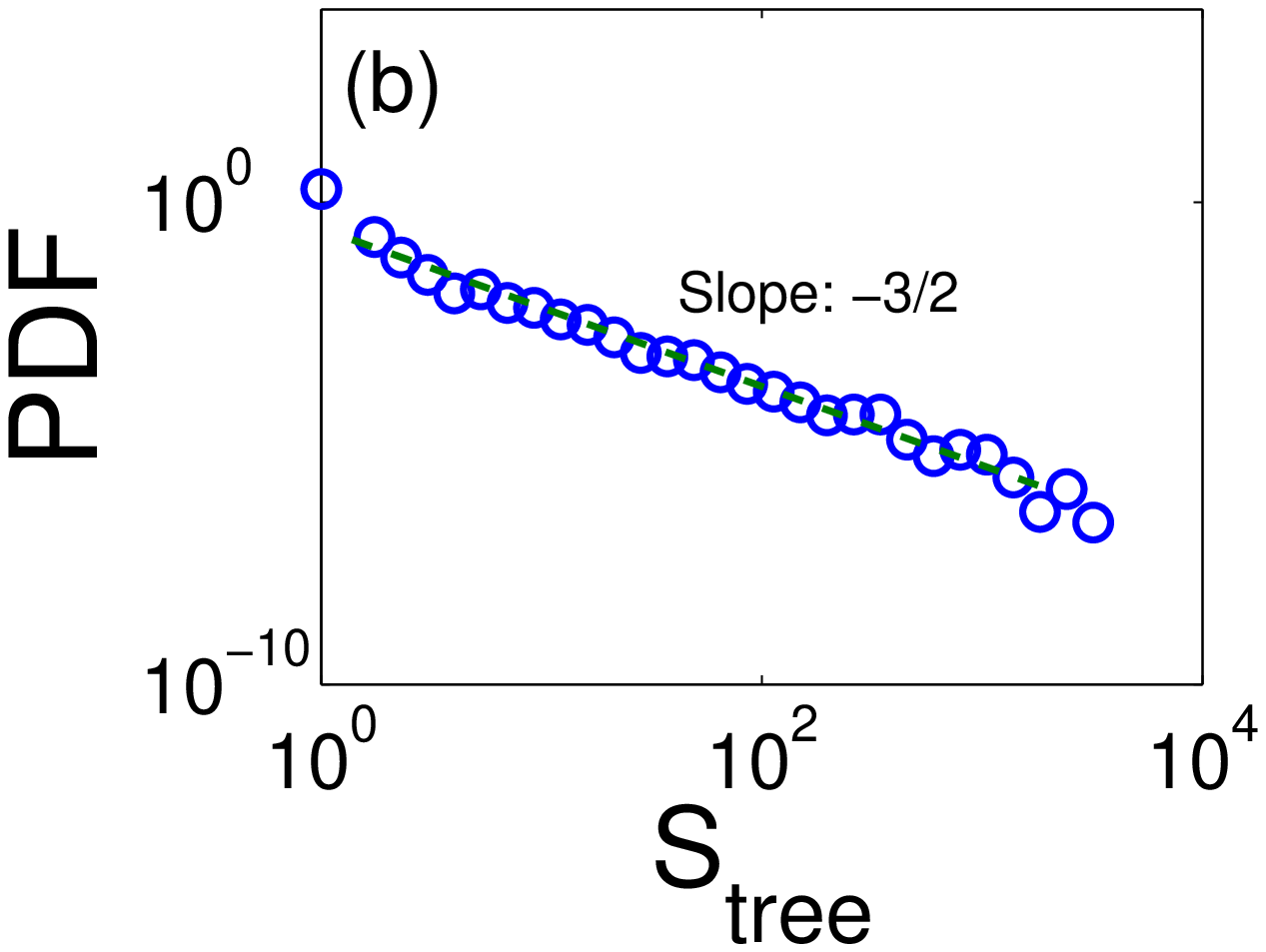}}
 \caption{(Color online) \textbf{a}. Scaling results of the mean and
 the standard deviation of the average failure size $\langle s_{t}\rangle$ from $T_{1}$
 to $T_{2}$, which is also the approximate number of branching processes for $p=p_{c}$ and
 $k=5$. Number or realizations is same as in Fig. \ref{fig2:subfig:b}. \textbf{b}. PDF of failure tree
 sizes $S_{tree}$ for the case $p=p_{c}$, $k=3$, $N=100\,000$ and 4537 trees in 80 realizations.}
 \label{fig4}
\end{figure}

Fig. \ref{fig3:subfig:a} and Fig. \ref{fig3:subfig:b} show the variation
of $s_{t}$ and $\eta_{t}$ respectively in a typical realization
that finally reached a total collapse. We observe that $\eta_{t}$
increases {during the cascades} from below $1$ to around $1$ (with some fluctuations) at the plateau, and
finally to above $1$ when the system starts to collapse. {The value of
$\eta_t$ is smaller than $1$ in the beginning of the cascading process since the
individual networks are still well connected and a large damage in one network leads to a smaller
damage in the second network (see Fig. \ref{fig3:subfig:a}). As cascading progresses the value of $\eta_t$
increases since both networks become more dilute and a failure in one step leads to relatively
higher damage in the next step (see Fig. \ref{fig3:subfig:b}). In this process the spontaneous
behavior of $\eta_t$ generates a new phase transition. When $\eta_t$ approaches
$1$ the system \textit{spontaneously enters} a second order percolation transition
where the cascading trees become critical branching processes (\cite{BUNbook}) of typical length
of $N^{1/3}$ as explained below. These long trees are the origin of the long plateau
observed in Fig. \ref{fig2:subfig:a}.}

The plateau stage starts when each of the $n$ failed nodes at iteration $T_1$ leads,
in average (we refer to the fluctuations explicitly in the following), to failure of
another single node (see \footnote{In order to estimate the length of the plateau
stage, we introduce a method to define the beginning, $T_1$, and the end, $T_2$, of
the plateau in each realization. In Fig. \ref{fig3:subfig:a}, we find the time step
$T_{f}$ of the first local minimum and $T_{l}$ of the last local minimum. Then,
we define a threshold $d=2\cdot\frac{1}{T_{l}-T_{f}+1}\cdot\sum\limits_{t=T_{f}}^{T_{l}}s_{t}$,
which is twice the average failure size between these two minimums. This is
because the $s_t$ values always have some random fluctuations above or below the mean value, which
should be the same order as the mean. Therefore, we use twice the mean as the threshold
for including such fluctuations. Then, we define $T_{1}$ and $T_{2}$ as the first time step and
the last one where $s_{t}\leq d$.}). This is a stable state, leading to the divergence of $\tau$
for $N\rightarrow\infty$. In a finite system of size $N$, however, the accumulated failures
slightly reduce $p$ and the number of failures at each iteration gradually increases. This bias
can be estimated by considering the percolation on single networks as follow.

At each time step $t$, the giant component size $\psi_{t}$ of network $A$ can be equivalently
regarded as randomly attacking a fraction $1-p$ on a single ER network. This specific
value of $p$, called the effective $p$ and denoted here by $p_{eff}$, can be obtained
theoretically by solving the equation $\psi_{t}=p\cdot g(p)${, where $g(p)$ is
the fraction of nodes in the giant component after randomly removing a fraction $p$ of nodes} (see \cite{BUL10A}).

Moreover, $\eta_t$ can be related to the branching factor for a single network.
Consider randomly removing a fraction $1-p$ of nodes in an ER network, which
makes some other nodes nonfunctional due to percolation, i.e., being disconnected
from the giant component. Then, we randomly remove one more node within the
giant component, and we use $\eta_{single}$ to denote the number of nodes that fail
additionally due to percolation. Notice that $\eta_{single}$ is the branching factor
for the additionally-removed node. Fig. \ref{fig3:subfig:c} shows the relation between $p$ and
$\eta_{single}$ for an ER network. Note that the branching factor diverges (for
infinite systems) when $p\rightarrow p_{c}^{+}$, and converges to $0$ when
$p\rightarrow 1$. Let $\tilde{p}$ be the critical value of $p$ where $\langle\eta_{single}\rangle=1$.
Then we see from Fig. \ref{fig3:subfig:c} that $\tilde{p}\approx 0.35$.

For two coupled ER networks, at each time step $t$ in the plateau stage,
the difference between the giant components of networks $A$ and $B$ is small
compared to the giant component sizes. Thus, each $A$($B$)-node that fails due
to dependency can be approximately regarded as randomly removing one more node
from the giant component of network $A$ ($B$). Therefore, $\eta_{t}\approx\langle\eta_{single}\rangle^{2}$
for the plateau stage. Notice that when $\langle\eta_{single}\rangle=1$,
$\eta_{t}$ also equals to 1, and the threshold $\tilde{p}\approx 0.35$
is also valid for coupled ER networks. This can be seen in Fig. \ref{fig3:subfig:d},
which shows the evolution of $p_{eff}$ in the same realization of Fig. \ref{fig3}.
We can see that the interaction between $p_{eff}$ and $\langle\eta_{t}\rangle$ is a
determinate factor for the plateau stage. As shown in Fig. \ref{fig3},
when $p_{eff}$ gets smaller, $\eta_{t}$ increases to about $1$. This causes a range of time
steps where $s_{t}$ is approximately a constant with some random fluctuations. Here, the random
fluctuations of $\eta_{t}$ will determine the end of the cascading process, with or without
a total collapse.

Based on these observations, we assume a random process of cascading failures starting at the
beginning of the plateau state at $t=T_{1}$. Let $n=s_{T_{1}}$, which is also the number of independent
failure trees, and consider time steps $T=t-T_{1}$. The variation of the failure sizes $s_{T}$ are determined by
both the systematic bias and the random fluctuations. Here, the random fluctuations can be described
by a Gaussian random walk from the value of $n$.

Assuming that $p_{eff}=\tilde{p}$, and $\eta_{T}=1$ at $T=0$, and $\eta_{T}$
decreases linearly when $p_{eff}$ increases near $\tilde{p}$: $\eta_{T}=1-C\cdot\Delta p_{eff}$.
Here, $C$ is a positive constant, and $\Delta p_{eff}$ is the increment of
$p_{eff}$ from $\tilde{p}$, which is approximately the variation of the giant
component size of network $A$. Therefore, $\Delta p_{eff}=-\frac{\sum\limits_{i=0}^{T}s_{i}}{N}$.
At $T=1$, we have $s_{1}=n\cdot(1-C\Delta p_{eff})=n\cdot(1+C\frac{n}{N})=n+\frac{C}{N}n^{2}$.
At $T=2$, we have $s_{2}=s_{1}\cdot(1+\frac{C}{N}(n+s_{1}))$.
After casting down small terms, we obtain $s_{2}=n+3\frac{C}{N}n^{2}$.
Similarly, we can obtain at $T$:

\begin{equation}
 s_{T}=n+\left(\sum\limits_{i=1}^{T}i\right)\cdot\frac{C}{N}n^{2}=n+\frac{T(T+1)}{2}\cdot\frac{C}{N}n^{2}.
\end{equation}

\noindent Therefore, the order of the systematic bias of failure sizes from $T_{1}$ to $T_{2}$
is $\frac{n^{2}T^{2}}{N}$. If at some iteration the number of failures becomes zero the cascading
process stops and the system survives. This can happen when $n-\sqrt{n}\sqrt{T}=0$, thus,

\begin{equation}
\label{eq8}
 T_{stop} \sim \sqrt{n}.
\end{equation}

\noindent If it does not stop, the cascading failures continue, and for large $T$
the bias will grow (faster than the fluctuations) leading to complete collapse.
The balance between the bias and the fluctuations may continue as long as

\begin{equation}
\label{eq9}
 \frac{n^{2}T^{2}}{N}\sim\sqrt{n}\sqrt{T}.
\end{equation}

\noindent Equations (\ref{eq8}) and (\ref{eq9}) yield that $n\sim T\sim N^{1/3}$,
which is supported by our simulation results in Fig. \ref{fig2:subfig:b} showing
$\langle\tau\rangle\equiv T\sim N^{1/3}$.

The above analysis also leads to the scaling law for the failure size at the beginning of the
plateau stage: $n\sim N^{1/3}$. This is supported by simulations shown in Fig. \ref{fig4:subfig:a},
which exhibits the average failure size $\langle s_{t}\rangle$ along the plateau stage near criticality.

The critical behavior at the plateau is also represented in the distribution
of failure tree sizes obtained in simulations shown in Fig. \ref{fig4:subfig:b}.
Here, we determine the beginning and the end of the plateau (see \footnotemark[2]),
and identify all $A$-nodes that fail due to dependency
in each of the parallel failure trees. At each time
step, the growth of each tree is determined by the branching factor $\eta_{t}$.
On the plateau, most trees will rapidly die out, while several trees keep growing and
become large. Fig. \ref{fig4:subfig:b} displays the PDF of the tree size $S_{tree}$, which
is the total number of nodes on a failure tree from the root to the time step where
it terminates. We can see that the total tree size has a power-law distribution with
a slope of approximately $-3/2$. It is interesting to note that such a distribution is
associated with cluster size distributions in second order percolation transitions,
see e.g., \cite{BUNbook,STAbook,COHbook} and obtained in classical models of self-organized criticality
\cite{CAR04B,CHE05,NED06,DOB06,DOB04}. Notice also that the same critical exponent has
been observed in real data \cite{CAR04A,HOL06,BAK06,ANC05}.

\begin{figure}
 \centering
 \subfigure{
  \label{fig5:subfig:a}
  \includegraphics[width=0.23\textwidth]{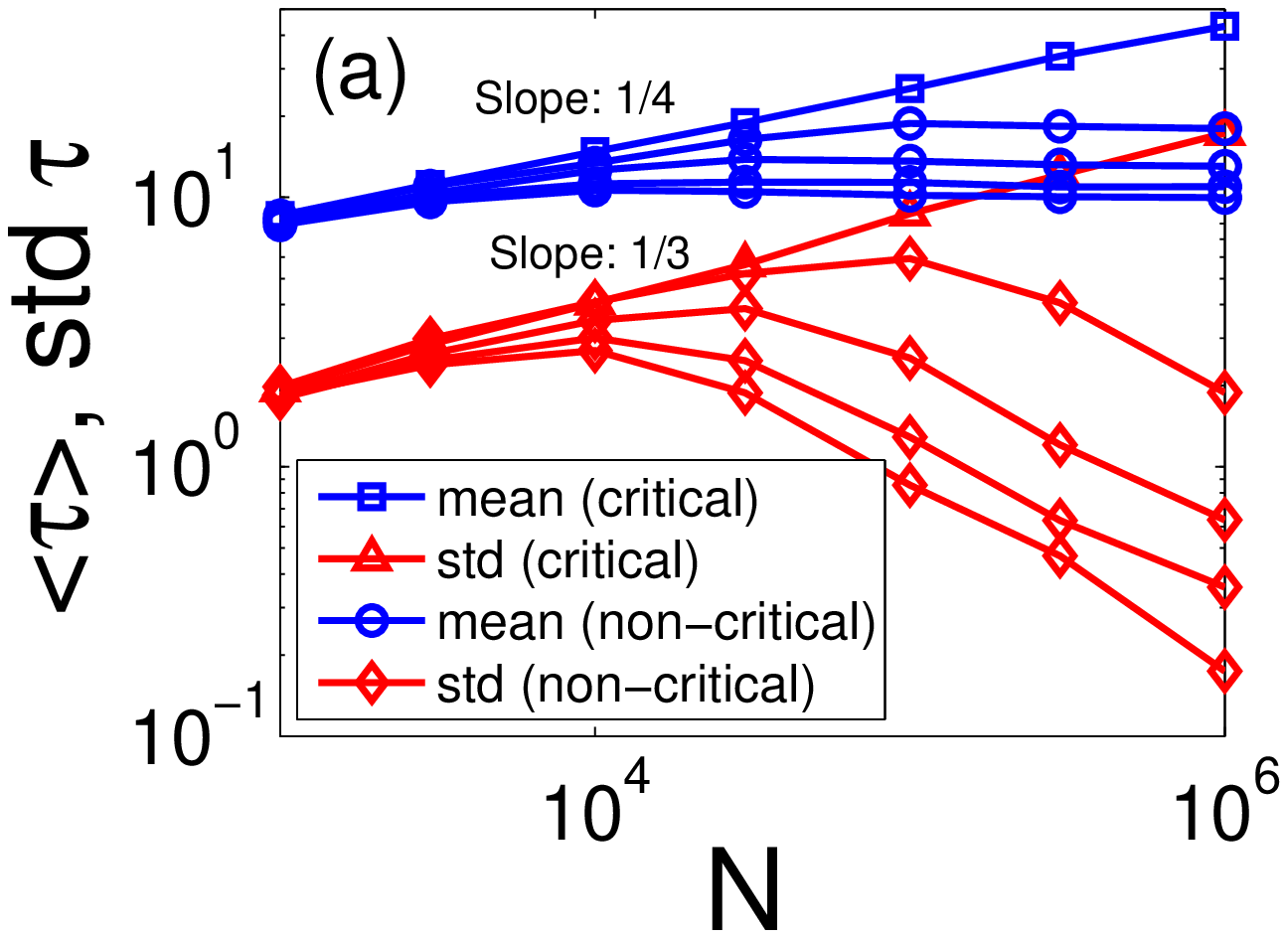}}
  \subfigure{
  \label{fig5:subfig:b}
  \includegraphics[width=0.23\textwidth]{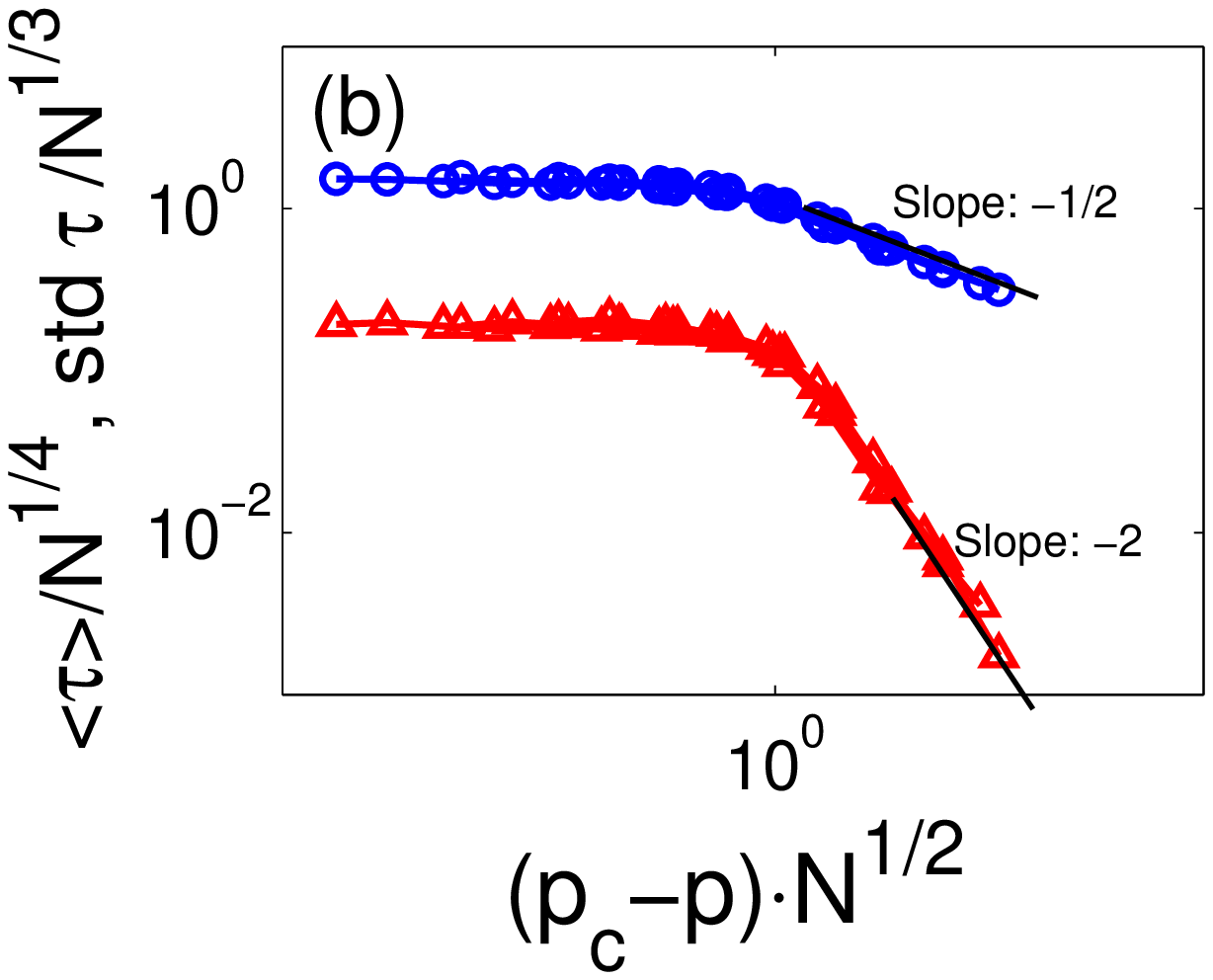}}
 \caption{(Color online) \textbf{a}. Scaling behaviors of the mean (blue)
 and the standard deviation (red) of the total time $\tau$ at $p^{MF}_{c}=2.4554/k$
 (critical) or below $p^{MF}_{c}$ (non-critical). Here,
 we consider $k=5$, and the number of realizations is $M=3000$. \textbf{b}.
 Scaled version of (a). Two more values of $p$ are included: $p=0.4908$ and
 $p=0.491$.}
 \label{fig5}
\end{figure}

\begin{figure}
 \centering
 \subfigure{
  \label{fig6:subfig:a}
  \includegraphics[width=0.23\textwidth]{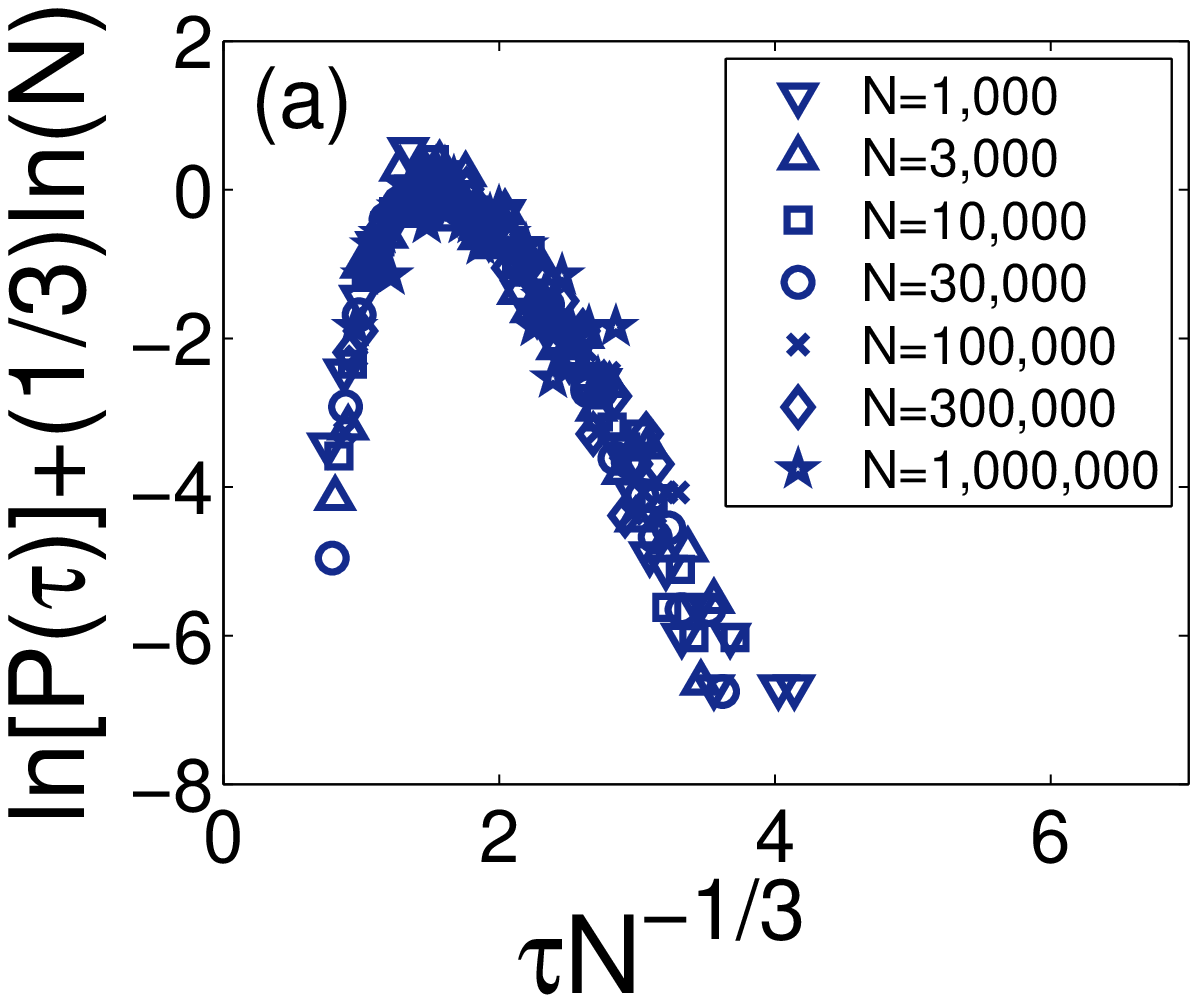}}
 \subfigure{
  \label{fig6:subfig:b}
  \includegraphics[width=0.23\textwidth]{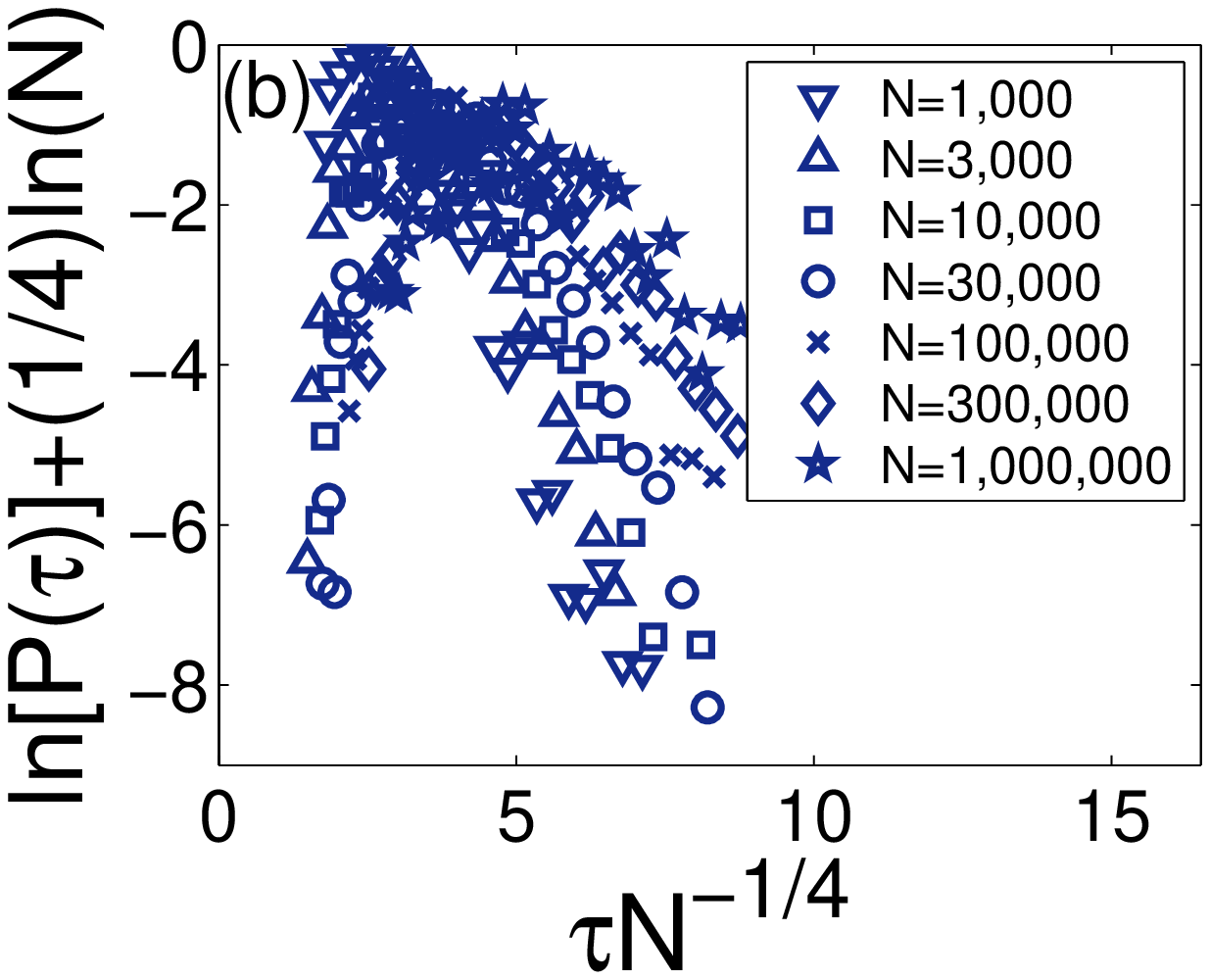}}
 \caption{(Color online) \textbf{a}. Scaled PDF of $\tau$ using the exponent $1/3$ at
 $p=p_{c}$ {for single realizations}. Here, $k=5$, and the number of realizations is the same as
 in Fig. 2(b). \textbf{b}. Scaled PDF of $\tau$ using the exponent $1/4$.}
 \label{fig6}
\end{figure}

\section{Relation between the Critical Scaling and the Mean-Field Case}

Buldyrev \textit{et al}. \cite{BUL10A} found both analytically and numerically
the scaling behavior $\langle\tau\rangle\sim N^{1/4}$ at $p^{MF}_{c}$,
which is significantly different from the critical scaling result
$N^{1/3}$ found here at $p_{c}$ of each realization. Fig. \ref{fig5} shows
the scaling behaviors of {both} $\langle\tau\rangle$ and
$\textup{\textbf{std}}(\tau)$ at and below $p^{MF}_{c}$. As can be seen, the
mean behavior is indeed consistent with the MF predictions of \cite{BUL10A}.
We will explain {in this section} this seemingly discrepancy by analyzing
the theoretical relationship between the scaling behaviors at $p_{c}$
of single realizations and at the MF prediction $p^{MF}_{c}$.

In Fig. \ref{fig5}, we observe the scaling rule of $\langle\tau\rangle$:
$\langle\tau\rangle\sim N^{1/4}\cdot f(u)$, where, $u=(p^{MF}_{c}-p)\cdot N^{1/\alpha}$,
and $\alpha=2$. Then, we have $f(u)\sim 1$ for $u\ll1$, and $f(u)\sim u^{-1/2}$
for $u\gg1$. Finally, for $N<N^{\ast}\sim(p_{c}-p)^{-3/2}$, $\langle\tau\rangle\sim N^{1/4}$,
and for $N>N^{\ast}$, $\langle\tau\rangle\sim (p_{c}-p)^{-1/2}$. Compared with
the scaling results for single realizations, Eqs. (\ref{eq1}) and (\ref{eq2}), the
main difference is the exponent $1/4$ of {the scaling of $\langle\tau\rangle$ with $N$. To further
validate our new scaling law, $\langle\tau\rangle\sim N^{1/3}$ at $p=p_c$ of single realizations,
we also compare in Fig. \ref{fig6} the two scaling laws for the PDF of $\tau$: Fig. \ref{fig6:subfig:a}
presents the PDF of $\tau$ for different values of $N$ according to the scaling assumption
$\tau\sim N^{1/3}$, whereas Fig. \ref{fig6:subfig:b} gives the PDF of $\tau$ according to the scaling
assumption $\tau\sim N^{1/4}$. As can bee seen from these figures, the assumption
$\tau\sim N^{1/3}$ seems to better fit the scaling for single realizations, further
supporting Eq. (\ref{eq2}).}

The origin of the MF observation, {$\langle\tau\rangle\sim N^{1/4}$ and
$\textup{\textbf{std}}(\tau)\sim N^{1/3}$ (see Fig. \ref{fig5}), which deviates from
Eq. (\ref{eq2}) for single realizations, can be explained by considering the fluctuations
which do not appear in the MF case.}

Given $\langle\tau\rangle\sim N^{1/3}$ at $p_{c}$, and $\langle\tau\rangle\sim (p_{c}-p)^{-1/2}$ when
$p$ is below $p_{c}$, the scaling behavior at $p^{MF}_{c}$ can be regarded as the expectation of
$\langle\tau\rangle$ below $p_{c}$:

\begin{equation}
 \langle\tau\rangle_{MF}=\int^{\infty}_{0}\langle\tau\rangle\cdot D(x)dx,
\end{equation}

where $x=p_{c}-p$, and $D(x)$ is its probability density. From the
scaling results in Fig. \ref{fig2:subfig:c}, we know that $\langle\tau\rangle\sim N^{1/3}$,
for $p_{c}-p<N^{-2/3}$ and $\langle\tau\rangle\sim (p_{c}-p)^{-1/2}$, for $p_{c}-p>N^{-2/3}$.
We also assume that the value of $p_{c}$ follows a Gaussian distribution
$N(p^{MF}_{c},\sigma^2)$ (This is supported by the distribution of $p_c$ in
simulations, see Fig. \ref{fig7} in the Appendix.) above $p^{MF}_{c}$, where
$\sigma\sim N^{-1/2}$. Therefore,

\begin{eqnarray}
\label{eqAPP1}
 \langle\tau\rangle_{MF}=I_{1}+I_{2}\sim\int^{N^{-2/3}}_{0}N^{1/3}\cdot\frac{1}{\sqrt{2\pi}\sigma}\exp\left(-\frac{x^2}{2\sigma^2}\right)dx
 \nonumber\\
+ \int^{\infty}_{N^{-2/3}}x^{-1/2}\cdot\frac{1}{\sqrt{2\pi}\sigma}\exp\left(-\frac{x^2}{2\sigma^2}\right)dx.
\end{eqnarray}

\noindent Let $y=x\cdot\sqrt{N}$, which finally yields $I_1\sim N^{1/6}$ and $I_2\sim N^{1/4}$, from
which follows $\langle\tau\rangle_{MF}=I_{1}+I_{2}\sim N^{1/4}$ for large $N$.

Similarly, we can also calculate the variance of $\tau$ using $\textup{\textbf{var}}(\tau)=\langle\tau^{2}\rangle-\langle\tau\rangle^{2}$,
and then estimate the scaling result for the standard deviation at $p^{MF}_{c}$. We can finally
obtain $\textup{\textbf{std}}(\tau)\sim N^{1/4}$, instead of $N^{1/3}$, seen in Fig. \ref{fig5:subfig:a}.
The explanation of this deviation can be understood by performing accurate numerical integrals for the
analogous Eq. (\ref{eqAPP1}) for the standard deviation. This accurate integration shows
that for small values of $N$, the scaling of $\textup{\textbf{std}}(\tau)$ with $N$ can be approximated
as $\textup{\textbf{std}}(\tau)\sim N^{1/3}$. However, for large $N$, the slope decreases to $N^{1/4}$.
This might explain for the slope $1/3$ of $\textup{\textbf{std}}(\tau)$ at $p^{MF}_{c}$ observed in
simulations, as shown in Fig. \ref{fig5:subfig:a}.

\

\section{Summary}

In this paper, we identified a spontaneous second order percolation transition occurring
during the cascading failures which controls the first order abrupt transition.
This {spontaneous transition} is characterized by {cascading of failure trees whose} size distribution is power law with an exponent $-3/2$ during the plateau stage. This explains
the origin of the long plateau and its scaling with $N$ found in cascading process near
the abrupt collapse of the coupled networks system. We also uncovered the theoretical
relationship between the two seemingly contradictory scaling, $\langle\tau\rangle\sim N^{1/4}$ at
the mean-field criticality and $\langle\tau\rangle\sim N^{1/3}$ at $p_c$ {of single realizations}, by considering the
deviation of $p_c$ in different realizations.

\section*{Acknowledgments}
We thank DTRA, ONR, {BSF,} the LINC (No. 289447) and the Multiplex (No. 317532) EU projects, the DFG, and
the Israel Science Foundation for support. {We thank Michael M. Danziger for
helpful discussions.}
\bibliography{manuscript}

\begin{thebibliography}{38}%
\makeatletter
\providecommand \@ifxundefined [1]{%
 \@ifx{#1\undefined}
}%
\providecommand \@ifnum [1]{%
 \ifnum #1\expandafter \@firstoftwo
 \else \expandafter \@secondoftwo
 \fi
}%
\providecommand \@ifx [1]{%
 \ifx #1\expandafter \@firstoftwo
 \else \expandafter \@secondoftwo
 \fi
}%
\providecommand \natexlab [1]{#1}%
\providecommand \enquote  [1]{``#1''}%
\providecommand \bibnamefont  [1]{#1}%
\providecommand \bibfnamefont [1]{#1}%
\providecommand \citenamefont [1]{#1}%
\providecommand \href@noop [0]{\@secondoftwo}%
\providecommand \href [0]{\begingroup \@sanitize@url \@href}%
\providecommand \@href[1]{\@@startlink{#1}\@@href}%
\providecommand \@@href[1]{\endgroup#1\@@endlink}%
\providecommand \@sanitize@url [0]{\catcode `\\12\catcode `\$12\catcode
  `\&12\catcode `\#12\catcode `\^12\catcode `\_12\catcode `\%12\relax}%
\providecommand \@@startlink[1]{}%
\providecommand \@@endlink[0]{}%
\providecommand \url  [0]{\begingroup\@sanitize@url \@url }%
\providecommand \@url [1]{\endgroup\@href {#1}{\urlprefix }}%
\providecommand \urlprefix  [0]{URL }%
\providecommand \Eprint [0]{\href }%
\providecommand \doibase [0]{http://dx.doi.org/}%
\providecommand \selectlanguage [0]{\@gobble}%
\providecommand \bibinfo  [0]{\@secondoftwo}%
\providecommand \bibfield  [0]{\@secondoftwo}%
\providecommand \translation [1]{[#1]}%
\providecommand \BibitemOpen [0]{}%
\providecommand \bibitemStop [0]{}%
\providecommand \bibitemNoStop [0]{.\EOS\space}%
\providecommand \EOS [0]{\spacefactor3000\relax}%
\providecommand \BibitemShut  [1]{\csname bibitem#1\endcsname}%
\let\auto@bib@innerbib\@empty
\bibitem [{\citenamefont {Laprie}\ \emph {et~al.}(2007)\citenamefont {Laprie},
  \citenamefont {Kanoun},\ and\ \citenamefont {Kaniche}}]{LAP07}%
  \BibitemOpen
  \bibfield  {author} {\bibinfo {author} {\bibfnamefont {J.}~\bibnamefont
  {Laprie}}, \bibinfo {author} {\bibfnamefont {K.}~\bibnamefont {Kanoun}}, \
  and\ \bibinfo {author} {\bibfnamefont {M.}~\bibnamefont {Kaniche}},\
  }\href@noop {} {\bibfield  {journal} {\bibinfo  {journal} {Lect. Notes
  Comput. Sci.}\ }\textbf {\bibinfo {volume} {4680}},\ \bibinfo {pages} {54}
  (\bibinfo {year} {2007})}\BibitemShut {NoStop}%
\bibitem [{\citenamefont {Rosato}\ \emph {et~al.}(2008)\citenamefont {Rosato},
  \citenamefont {Issacharoff}, \citenamefont {Tiriticco}, \citenamefont
  {Meloni}, \citenamefont {De~Porcellinis},\ and\ \citenamefont
  {Setola}}]{ROS08}%
  \BibitemOpen
  \bibfield  {author} {\bibinfo {author} {\bibfnamefont {V.}~\bibnamefont
  {Rosato}}, \bibinfo {author} {\bibfnamefont {L.}~\bibnamefont {Issacharoff}},
  \bibinfo {author} {\bibfnamefont {F.}~\bibnamefont {Tiriticco}}, \bibinfo
  {author} {\bibfnamefont {S.}~\bibnamefont {Meloni}}, \bibinfo {author}
  {\bibfnamefont {S.}~\bibnamefont {De~Porcellinis}}, \ and\ \bibinfo {author}
  {\bibfnamefont {R.}~\bibnamefont {Setola}},\ }\href@noop {} {\bibfield
  {journal} {\bibinfo  {journal} {Int. J. Crit. Infrastruct}\ }\textbf
  {\bibinfo {volume} {4}},\ \bibinfo {pages} {63} (\bibinfo {year}
  {2008})}\BibitemShut {NoStop}%
\bibitem [{\citenamefont {Panzieri}\ and\ \citenamefont
  {Setola}(2008)}]{PAN08}%
  \BibitemOpen
  \bibfield  {author} {\bibinfo {author} {\bibfnamefont {S.}~\bibnamefont
  {Panzieri}}\ and\ \bibinfo {author} {\bibfnamefont {R.}~\bibnamefont
  {Setola}},\ }\href@noop {} {\bibfield  {journal} {\bibinfo  {journal} {Int.
  J. Model. Ident. Contr.}\ }\textbf {\bibinfo {volume} {3}},\ \bibinfo {pages}
  {69} (\bibinfo {year} {2008})}\BibitemShut {NoStop}%
\bibitem [{\citenamefont {Vespignani}(2010)}]{VES10}%
  \BibitemOpen
  \bibfield  {author} {\bibinfo {author} {\bibfnamefont {A.}~\bibnamefont
  {Vespignani}},\ }\href@noop {} {\bibfield  {journal} {\bibinfo  {journal}
  {Nature (London)}\ }\textbf {\bibinfo {volume} {464}},\ \bibinfo {pages}
  {984} (\bibinfo {year} {2010})}\BibitemShut {NoStop}%
\bibitem [{\citenamefont {Buldyrev}\ \emph {et~al.}(2010)\citenamefont
  {Buldyrev}, \citenamefont {Parshani}, \citenamefont {Paul}, \citenamefont
  {Stanley},\ and\ \citenamefont {Havlin}}]{BUL10A}%
  \BibitemOpen
  \bibfield  {author} {\bibinfo {author} {\bibfnamefont {S.~V.}\ \bibnamefont
  {Buldyrev}}, \bibinfo {author} {\bibfnamefont {R.}~\bibnamefont {Parshani}},
  \bibinfo {author} {\bibfnamefont {G.}~\bibnamefont {Paul}}, \bibinfo {author}
  {\bibfnamefont {H.~E.}\ \bibnamefont {Stanley}}, \ and\ \bibinfo {author}
  {\bibfnamefont {S.}~\bibnamefont {Havlin}},\ }\href@noop {} {\bibfield
  {journal} {\bibinfo  {journal} {Nature (London)}\ }\textbf {\bibinfo {volume}
  {464}},\ \bibinfo {pages} {1025} (\bibinfo {year} {2010})}\BibitemShut
  {NoStop}%
\bibitem [{\citenamefont {Parshani}\ \emph
  {et~al.}(2010{\natexlab{a}})\citenamefont {Parshani}, \citenamefont
  {Buldyrev},\ and\ \citenamefont {Havlin}}]{PAR10A}%
  \BibitemOpen
  \bibfield  {author} {\bibinfo {author} {\bibfnamefont {R.}~\bibnamefont
  {Parshani}}, \bibinfo {author} {\bibfnamefont {S.~V.}\ \bibnamefont
  {Buldyrev}}, \ and\ \bibinfo {author} {\bibfnamefont {S.}~\bibnamefont
  {Havlin}},\ }\href@noop {} {\bibfield  {journal} {\bibinfo  {journal} {Phys.
  Rev. Lett.}\ }\textbf {\bibinfo {volume} {105}},\ \bibinfo {pages} {048701}
  (\bibinfo {year} {2010}{\natexlab{a}})}\BibitemShut {NoStop}%
\bibitem [{\citenamefont {Gao}\ \emph {et~al.}(2011{\natexlab{a}})\citenamefont
  {Gao}, \citenamefont {Buldyrev}, \citenamefont {Stanley},\ and\ \citenamefont
  {Havlin}}]{GAO11}%
  \BibitemOpen
  \bibfield  {author} {\bibinfo {author} {\bibfnamefont {J.}~\bibnamefont
  {Gao}}, \bibinfo {author} {\bibfnamefont {S.~V.}\ \bibnamefont {Buldyrev}},
  \bibinfo {author} {\bibfnamefont {H.~E.}\ \bibnamefont {Stanley}}, \ and\
  \bibinfo {author} {\bibfnamefont {S.}~\bibnamefont {Havlin}},\ }\href@noop {}
  {\bibfield  {journal} {\bibinfo  {journal} {Nature Physics}\ }\textbf
  {\bibinfo {volume} {8}},\ \bibinfo {pages} {40} (\bibinfo {year}
  {2011}{\natexlab{a}})}\BibitemShut {NoStop}%
\bibitem [{\citenamefont {Leicht}\ and\ \citenamefont {D'Souza}(2009)}]{LEI09}%
  \BibitemOpen
  \bibfield  {author} {\bibinfo {author} {\bibfnamefont {E.~A.}\ \bibnamefont
  {Leicht}}\ and\ \bibinfo {author} {\bibfnamefont {R.~M.}\ \bibnamefont
  {D'Souza}},\ }\href@noop {} {\enquote {\bibinfo {title} {Percolation on
  interacting networks},}\ } (\bibinfo {year} {2009}),\ \Eprint
  {http://arxiv.org/abs/0907.0894} {arXiv:0907.0894} \BibitemShut {NoStop}%
\bibitem [{\citenamefont {Morris}\ and\ \citenamefont
  {Barthelemy}(2012)}]{MOR12}%
  \BibitemOpen
  \bibfield  {author} {\bibinfo {author} {\bibfnamefont {R.~G.}\ \bibnamefont
  {Morris}}\ and\ \bibinfo {author} {\bibfnamefont {M.}~\bibnamefont
  {Barthelemy}},\ }\href@noop {} {\bibfield  {journal} {\bibinfo  {journal}
  {Phys. Rev. Lett.}\ }\textbf {\bibinfo {volume} {109}},\ \bibinfo {pages}
  {128703} (\bibinfo {year} {2012})}\BibitemShut {NoStop}%
\bibitem [{\citenamefont {Saumell-Mendiola}\ \emph {et~al.}(2012)\citenamefont
  {Saumell-Mendiola}, \citenamefont {Serrano},\ and\ \citenamefont
  {Bogu\~n\'a}}]{SAU12}%
  \BibitemOpen
  \bibfield  {author} {\bibinfo {author} {\bibfnamefont {A.}~\bibnamefont
  {Saumell-Mendiola}}, \bibinfo {author} {\bibfnamefont {M.~A.}\ \bibnamefont
  {Serrano}}, \ and\ \bibinfo {author} {\bibfnamefont {M.}~\bibnamefont
  {Bogu\~n\'a}},\ }\href@noop {} {\bibfield  {journal} {\bibinfo  {journal}
  {Phys. Rev. E}\ }\textbf {\bibinfo {volume} {86}},\ \bibinfo {pages} {026106}
  (\bibinfo {year} {2012})}\BibitemShut {NoStop}%
\bibitem [{\citenamefont {Shao}\ \emph {et~al.}(2011)\citenamefont {Shao},
  \citenamefont {Buldyrev}, \citenamefont {Havlin},\ and\ \citenamefont
  {Stanley}}]{SHA11}%
  \BibitemOpen
  \bibfield  {author} {\bibinfo {author} {\bibfnamefont {J.}~\bibnamefont
  {Shao}}, \bibinfo {author} {\bibfnamefont {S.~V.}\ \bibnamefont {Buldyrev}},
  \bibinfo {author} {\bibfnamefont {S.}~\bibnamefont {Havlin}}, \ and\ \bibinfo
  {author} {\bibfnamefont {H.~E.}\ \bibnamefont {Stanley}},\ }\href@noop {}
  {\bibfield  {journal} {\bibinfo  {journal} {Phys. Rev. E}\ }\textbf {\bibinfo
  {volume} {83}},\ \bibinfo {pages} {036116} (\bibinfo {year}
  {2011})}\BibitemShut {NoStop}%
\bibitem [{\citenamefont {Gao}\ \emph {et~al.}(2011{\natexlab{b}})\citenamefont
  {Gao}, \citenamefont {Buldyrev}, \citenamefont {Havlin},\ and\ \citenamefont
  {Stanley}}]{GAO10}%
  \BibitemOpen
  \bibfield  {author} {\bibinfo {author} {\bibfnamefont {J.}~\bibnamefont
  {Gao}}, \bibinfo {author} {\bibfnamefont {S.~V.}\ \bibnamefont {Buldyrev}},
  \bibinfo {author} {\bibfnamefont {S.}~\bibnamefont {Havlin}}, \ and\ \bibinfo
  {author} {\bibfnamefont {H.~E.}\ \bibnamefont {Stanley}},\ }\href@noop {}
  {\bibfield  {journal} {\bibinfo  {journal} {Phys. Rev. Lett.}\ }\textbf
  {\bibinfo {volume} {107}},\ \bibinfo {pages} {195701} (\bibinfo {year}
  {2011}{\natexlab{b}})}\BibitemShut {NoStop}%
\bibitem [{\citenamefont {Huang}\ \emph {et~al.}(2011)\citenamefont {Huang},
  \citenamefont {Gao}, \citenamefont {Buldyrev}, \citenamefont {Havlin},\ and\
  \citenamefont {Stanley}}]{HUA11}%
  \BibitemOpen
  \bibfield  {author} {\bibinfo {author} {\bibfnamefont {X.}~\bibnamefont
  {Huang}}, \bibinfo {author} {\bibfnamefont {J.}~\bibnamefont {Gao}}, \bibinfo
  {author} {\bibfnamefont {S.~V.}\ \bibnamefont {Buldyrev}}, \bibinfo {author}
  {\bibfnamefont {S.}~\bibnamefont {Havlin}}, \ and\ \bibinfo {author}
  {\bibfnamefont {H.~E.}\ \bibnamefont {Stanley}},\ }\href@noop {} {\bibfield
  {journal} {\bibinfo  {journal} {Phys. Rev. E}\ }\textbf {\bibinfo {volume}
  {83}},\ \bibinfo {pages} {065101} (\bibinfo {year} {2011})}\BibitemShut
  {NoStop}%
\bibitem [{\citenamefont {G\'omez}\ \emph {et~al.}(2013)\citenamefont
  {G\'omez}, \citenamefont {D\'iaz-Guilera}, \citenamefont
  {G\'omez-Garde\~nes}, \citenamefont {P\'erez-Vicente}, \citenamefont
  {Moreno},\ and\ \citenamefont {Arenas}}]{GOM13}%
  \BibitemOpen
  \bibfield  {author} {\bibinfo {author} {\bibfnamefont {S.}~\bibnamefont
  {G\'omez}}, \bibinfo {author} {\bibfnamefont {A.}~\bibnamefont
  {D\'iaz-Guilera}}, \bibinfo {author} {\bibfnamefont {J.}~\bibnamefont
  {G\'omez-Garde\~nes}}, \bibinfo {author} {\bibfnamefont {C.~J.}\ \bibnamefont
  {P\'erez-Vicente}}, \bibinfo {author} {\bibfnamefont {Y.}~\bibnamefont
  {Moreno}}, \ and\ \bibinfo {author} {\bibfnamefont {A.}~\bibnamefont
  {Arenas}},\ }\href@noop {} {\bibfield  {journal} {\bibinfo  {journal} {Phys.
  Rev. Lett.}\ }\textbf {\bibinfo {volume} {110}},\ \bibinfo {pages} {028701}
  (\bibinfo {year} {2013})}\BibitemShut {NoStop}%
\bibitem [{\citenamefont {Aguirre}\ \emph {et~al.}(2013)\citenamefont
  {Aguirre}, \citenamefont {Papo},\ and\ \citenamefont {Buld\'u}}]{AGU13}%
  \BibitemOpen
  \bibfield  {author} {\bibinfo {author} {\bibfnamefont {J.}~\bibnamefont
  {Aguirre}}, \bibinfo {author} {\bibfnamefont {D.}~\bibnamefont {Papo}}, \
  and\ \bibinfo {author} {\bibfnamefont {J.~M.}\ \bibnamefont {Buld\'u}},\
  }\href@noop {} {\bibfield  {journal} {\bibinfo  {journal} {Nature Physics}\
  }\textbf {\bibinfo {volume} {9}},\ \bibinfo {pages} {230} (\bibinfo {year}
  {2013})}\BibitemShut {NoStop}%
\bibitem [{\citenamefont {Brummitt}\ and\ \citenamefont
  {R.~M.~D'Souza}(2012)}]{CHA12}%
  \BibitemOpen
  \bibfield  {author} {\bibinfo {author} {\bibfnamefont {C.~D.}\ \bibnamefont
  {Brummitt}}\ and\ \bibinfo {author} {\bibfnamefont {E.~A.~L.}\ \bibnamefont
  {R.~M.~D'Souza}},\ }\href@noop {} {\bibfield  {journal} {\bibinfo  {journal}
  {Proc. Natl. Acad. Sci.}\ }\textbf {\bibinfo {volume} {109}},\ \bibinfo
  {pages} {E680} (\bibinfo {year} {2012})}\BibitemShut {NoStop}%
\bibitem [{\citenamefont {Bianconi}(2013)}]{BIA13}%
  \BibitemOpen
  \bibfield  {author} {\bibinfo {author} {\bibfnamefont {G.}~\bibnamefont
  {Bianconi}},\ }\href@noop {} {\bibfield  {journal} {\bibinfo  {journal}
  {Phys. Rev. E}\ }\textbf {\bibinfo {volume} {87}},\ \bibinfo {pages} {062806}
  (\bibinfo {year} {2013})}\BibitemShut {NoStop}%
\bibitem [{\citenamefont {Cellai}\ \emph {et~al.}(2013)\citenamefont {Cellai},
  \citenamefont {L\'opez}, \citenamefont {Zhou}, \citenamefont {Gleeson},\ and\
  \citenamefont {Bianconi}}]{CEL13}%
  \BibitemOpen
  \bibfield  {author} {\bibinfo {author} {\bibfnamefont {D.}~\bibnamefont
  {Cellai}}, \bibinfo {author} {\bibfnamefont {E.}~\bibnamefont {L\'opez}},
  \bibinfo {author} {\bibfnamefont {J.}~\bibnamefont {Zhou}}, \bibinfo {author}
  {\bibfnamefont {J.~P.}\ \bibnamefont {Gleeson}}, \ and\ \bibinfo {author}
  {\bibfnamefont {G.}~\bibnamefont {Bianconi}},\ }\href@noop {} {\bibfield
  {journal} {\bibinfo  {journal} {Phys. Rev. E}\ }\textbf {\bibinfo {volume}
  {88}},\ \bibinfo {pages} {052811} (\bibinfo {year} {2013})}\BibitemShut
  {NoStop}%
\bibitem [{\citenamefont {Radicchi}\ and\ \citenamefont
  {Arenas}(2013)}]{RAD13}%
  \BibitemOpen
  \bibfield  {author} {\bibinfo {author} {\bibfnamefont {F.}~\bibnamefont
  {Radicchi}}\ and\ \bibinfo {author} {\bibfnamefont {A.}~\bibnamefont
  {Arenas}},\ }\href@noop {} {\bibfield  {journal} {\bibinfo  {journal} {Nature
  Physics}\ }\textbf {\bibinfo {volume} {9}},\ \bibinfo {pages} {717} (\bibinfo
  {year} {2013})}\BibitemShut {NoStop}%
\bibitem [{\citenamefont {Li}\ \emph {et~al.}(2012)\citenamefont {Li},
  \citenamefont {Bashan}, \citenamefont {Buldyrev}, \citenamefont {Stanley},\
  and\ \citenamefont {Havlin}}]{LI12}%
  \BibitemOpen
  \bibfield  {author} {\bibinfo {author} {\bibfnamefont {W.}~\bibnamefont
  {Li}}, \bibinfo {author} {\bibfnamefont {A.}~\bibnamefont {Bashan}}, \bibinfo
  {author} {\bibfnamefont {S.~V.}\ \bibnamefont {Buldyrev}}, \bibinfo {author}
  {\bibfnamefont {H.~E.}\ \bibnamefont {Stanley}}, \ and\ \bibinfo {author}
  {\bibfnamefont {S.}~\bibnamefont {Havlin}},\ }\href@noop {} {\bibfield
  {journal} {\bibinfo  {journal} {Phys. Rev. Lett.}\ }\textbf {\bibinfo
  {volume} {108}},\ \bibinfo {pages} {228702} (\bibinfo {year}
  {2012})}\BibitemShut {NoStop}%
\bibitem [{\citenamefont {Bashan}\ \emph {et~al.}(2013)\citenamefont {Bashan},
  \citenamefont {Berezin}, \citenamefont {Buldyrev},\ and\ \citenamefont
  {Havlin}}]{BAS13}%
  \BibitemOpen
  \bibfield  {author} {\bibinfo {author} {\bibfnamefont {A.}~\bibnamefont
  {Bashan}}, \bibinfo {author} {\bibfnamefont {Y.}~\bibnamefont {Berezin}},
  \bibinfo {author} {\bibfnamefont {S.~V.}\ \bibnamefont {Buldyrev}}, \ and\
  \bibinfo {author} {\bibfnamefont {S.}~\bibnamefont {Havlin}},\ }\href@noop {}
  {\bibfield  {journal} {\bibinfo  {journal} {Nature Physics}\ }\textbf
  {\bibinfo {volume} {9}},\ \bibinfo {pages} {667} (\bibinfo {year}
  {2013})}\BibitemShut {NoStop}%
\bibitem [{\citenamefont {Parshani}\ \emph {et~al.}(2011)\citenamefont
  {Parshani}, \citenamefont {Buldyrev},\ and\ \citenamefont {Havlin}}]{PAR11}%
  \BibitemOpen
  \bibfield  {author} {\bibinfo {author} {\bibfnamefont {R.}~\bibnamefont
  {Parshani}}, \bibinfo {author} {\bibfnamefont {S.~V.}\ \bibnamefont
  {Buldyrev}}, \ and\ \bibinfo {author} {\bibfnamefont {S.}~\bibnamefont
  {Havlin}},\ }\href@noop {} {\bibfield  {journal} {\bibinfo  {journal} {Proc.
  Natl. Acad. Sci.}\ }\textbf {\bibinfo {volume} {108}},\ \bibinfo {pages}
  {1007} (\bibinfo {year} {2011})}\BibitemShut {NoStop}%
\bibitem [{\citenamefont {Carreras}\ \emph
  {et~al.}(2004{\natexlab{a}})\citenamefont {Carreras}, \citenamefont {Newman},
  \citenamefont {Dobson},\ and\ \citenamefont {Poole}}]{CAR04A}%
  \BibitemOpen
  \bibfield  {author} {\bibinfo {author} {\bibfnamefont {B.~A.}\ \bibnamefont
  {Carreras}}, \bibinfo {author} {\bibfnamefont {D.~E.}\ \bibnamefont
  {Newman}}, \bibinfo {author} {\bibfnamefont {I.}~\bibnamefont {Dobson}}, \
  and\ \bibinfo {author} {\bibfnamefont {A.~B.}\ \bibnamefont {Poole}},\
  }\href@noop {} {\bibfield  {journal} {\bibinfo  {journal} {IEEE Trans.
  Circuits Syst., I: Fundam. Theory Appl.}\ }\textbf {\bibinfo {volume} {51}},\
  \bibinfo {pages} {1733} (\bibinfo {year} {2004}{\natexlab{a}})}\BibitemShut
  {NoStop}%
\bibitem [{\citenamefont {$\mathring{A}$. J.~Holmgren}\ and\ \citenamefont
  {Molin}(2006)}]{HOL06}%
  \BibitemOpen
  \bibfield  {author} {\bibinfo {author} {\bibnamefont {$\mathring{A}$.
  J.~Holmgren}}\ and\ \bibinfo {author} {\bibfnamefont {S.}~\bibnamefont
  {Molin}},\ }\href@noop {} {\bibfield  {journal} {\bibinfo  {journal} {J.
  Infrastruct. Syst.}\ }\textbf {\bibinfo {volume} {12}},\ \bibinfo {pages}
  {243} (\bibinfo {year} {2006})}\BibitemShut {NoStop}%
\bibitem [{\citenamefont {Bakke}\ \emph {et~al.}(2006)\citenamefont {Bakke},
  \citenamefont {Hansen},\ and\ \citenamefont {Kert\'esz}}]{BAK06}%
  \BibitemOpen
  \bibfield  {author} {\bibinfo {author} {\bibfnamefont {J.}~\bibnamefont
  {Bakke}}, \bibinfo {author} {\bibfnamefont {A.}~\bibnamefont {Hansen}}, \
  and\ \bibinfo {author} {\bibfnamefont {J.}~\bibnamefont {Kert\'esz}},\
  }\href@noop {} {\bibfield  {journal} {\bibinfo  {journal} {Europhys. Lett.}\
  }\textbf {\bibinfo {volume} {76}},\ \bibinfo {pages} {717} (\bibinfo {year}
  {2006})}\BibitemShut {NoStop}%
\bibitem [{\citenamefont {Ancell}\ \emph {et~al.}(2005)\citenamefont {Ancell},
  \citenamefont {Edwards},\ and\ \citenamefont {Krichtal}}]{ANC05}%
  \BibitemOpen
  \bibfield  {author} {\bibinfo {author} {\bibfnamefont {G.}~\bibnamefont
  {Ancell}}, \bibinfo {author} {\bibfnamefont {C.}~\bibnamefont {Edwards}}, \
  and\ \bibinfo {author} {\bibfnamefont {V.}~\bibnamefont {Krichtal}},\ }in\
  \href@noop {} {\emph {\bibinfo {booktitle} {Electricity Engineers Association
  2005 Conference: Implementing New Zealand¡¯s Energy Options}}}\ (\bibinfo
  {address} {Aukland, New Zealand},\ \bibinfo {year} {2005})\BibitemShut
  {NoStop}%
\bibitem [{\citenamefont {Cohen}\ and\ \citenamefont {Havlin}(2010)}]{COHbook}%
  \BibitemOpen
  \bibfield  {author} {\bibinfo {author} {\bibfnamefont {R.}~\bibnamefont
  {Cohen}}\ and\ \bibinfo {author} {\bibfnamefont {S.}~\bibnamefont {Havlin}},\
  }\href@noop {} {\emph {\bibinfo {title} {Complex Networks: Structure,
  Robustness and Function}}}\ (\bibinfo  {publisher} {Cambridge University
  Press},\ \bibinfo {year} {2010})\BibitemShut {NoStop}%
\bibitem [{\citenamefont {Parshani}\ \emph
  {et~al.}(2010{\natexlab{b}})\citenamefont {Parshani}, \citenamefont
  {Rozenblat}, \citenamefont {Ietri}, \citenamefont {Ducruet},\ and\
  \citenamefont {Havlin}}]{PAR10B}%
  \BibitemOpen
  \bibfield  {author} {\bibinfo {author} {\bibfnamefont {R.}~\bibnamefont
  {Parshani}}, \bibinfo {author} {\bibfnamefont {C.}~\bibnamefont {Rozenblat}},
  \bibinfo {author} {\bibfnamefont {D.}~\bibnamefont {Ietri}}, \bibinfo
  {author} {\bibfnamefont {C.}~\bibnamefont {Ducruet}}, \ and\ \bibinfo
  {author} {\bibfnamefont {S.}~\bibnamefont {Havlin}},\ }\href@noop {}
  {\bibfield  {journal} {\bibinfo  {journal} {Europhys. Lett.}\ }\textbf
  {\bibinfo {volume} {92}},\ \bibinfo {pages} {68002} (\bibinfo {year}
  {2010}{\natexlab{b}})}\BibitemShut {NoStop}%
\bibitem [{\citenamefont {Hu}\ \emph {et~al.}(2013)\citenamefont {Hu},
  \citenamefont {Zhou}, \citenamefont {Zhang}, \citenamefont {Han},
  \citenamefont {Rozenblat},\ and\ \citenamefont {Havlin}}]{HU13}%
  \BibitemOpen
  \bibfield  {author} {\bibinfo {author} {\bibfnamefont {Y.}~\bibnamefont
  {Hu}}, \bibinfo {author} {\bibfnamefont {D.}~\bibnamefont {Zhou}}, \bibinfo
  {author} {\bibfnamefont {R.}~\bibnamefont {Zhang}}, \bibinfo {author}
  {\bibfnamefont {Z.}~\bibnamefont {Han}}, \bibinfo {author} {\bibfnamefont
  {C.}~\bibnamefont {Rozenblat}}, \ and\ \bibinfo {author} {\bibfnamefont
  {S.}~\bibnamefont {Havlin}},\ }\href@noop {} {\bibfield  {journal} {\bibinfo
  {journal} {Phys. Rev. E}\ }\textbf {\bibinfo {volume} {88}},\ \bibinfo
  {pages} {052805} (\bibinfo {year} {2013})}\BibitemShut {NoStop}%
\bibitem [{Note1()}]{Note1}%
  \BibitemOpen
  \bibinfo {note} {We tested also by taking a few different realizations and
  keeping the network structures but changing only the attack order. The
  results have been found to be very similar, as seen in Fig. \ref
  {fig8}}\BibitemShut {NoStop}%
\bibitem [{\citenamefont {Bunde}\ and\ \citenamefont {Havlin}(1996)}]{BUNbook}%
  \BibitemOpen
  \bibfield  {author} {\bibinfo {author} {\bibfnamefont {A.}~\bibnamefont
  {Bunde}}\ and\ \bibinfo {author} {\bibfnamefont {S.}~\bibnamefont {Havlin}},\
  }\href@noop {} {\emph {\bibinfo {title} {Fractals and Disordered Systems}}}\
  (\bibinfo  {publisher} {Springer},\ \bibinfo {year} {1996})\BibitemShut
  {NoStop}%
\bibitem [{\citenamefont {Stauffer}\ and\ \citenamefont
  {Aharony}(1992)}]{STAbook}%
  \BibitemOpen
  \bibfield  {author} {\bibinfo {author} {\bibfnamefont {D.}~\bibnamefont
  {Stauffer}}\ and\ \bibinfo {author} {\bibfnamefont {A.}~\bibnamefont
  {Aharony}},\ }\href@noop {} {\emph {\bibinfo {title} {Introduction to
  Percolation Theory}}}\ (\bibinfo  {publisher} {Taylor \& Francis},\ \bibinfo
  {year} {1992})\BibitemShut {NoStop}%
\bibitem [{Note2()}]{Note2}%
  \BibitemOpen
  \bibinfo {note} {In order to estimate the length of the plateau stage, we
  introduce a method to define the beginning, $T_1$, and the end, $T_2$, of the
  plateau in each realization. In Fig. \ref {fig3:subfig:a}, we find the time
  step $T_{f}$ of the first local minimum and $T_{l}$ of the last local
  minimum. Then, we define a threshold $d=2\cdot {\begingroup 1\endgroup \over
  T_{l}-T_{f}+1}\cdot \sum \limits _{t=T_{f}}^{T_{l}}s_{t}$, which is twice the
  average failure size between these two minimums. This is because the $s_t$
  values always have some random fluctuations above or below the mean value,
  which should be the same order as the mean. Therefore, we use twice the mean
  as the threshold for including such fluctuations. Then, we define $T_{1}$ and
  $T_{2}$ as the first time step and the last one where $s_{t}\leq
  d$.}\BibitemShut {Stop}%
\bibitem [{\citenamefont {Carreras}\ \emph
  {et~al.}(2004{\natexlab{b}})\citenamefont {Carreras}, \citenamefont {Lynch},
  \citenamefont {Dobson},\ and\ \citenamefont {Newman}}]{CAR04B}%
  \BibitemOpen
  \bibfield  {author} {\bibinfo {author} {\bibfnamefont {B.~A.}\ \bibnamefont
  {Carreras}}, \bibinfo {author} {\bibfnamefont {V.~E.}\ \bibnamefont {Lynch}},
  \bibinfo {author} {\bibfnamefont {I.}~\bibnamefont {Dobson}}, \ and\ \bibinfo
  {author} {\bibfnamefont {D.~E.}\ \bibnamefont {Newman}},\ }\href@noop {}
  {\bibfield  {journal} {\bibinfo  {journal} {Chaos}\ }\textbf {\bibinfo
  {volume} {14}},\ \bibinfo {pages} {643} (\bibinfo {year}
  {2004}{\natexlab{b}})}\BibitemShut {NoStop}%
\bibitem [{\citenamefont {Chen}\ \emph {et~al.}(2005)\citenamefont {Chen},
  \citenamefont {Thorp},\ and\ \citenamefont {Dobson}}]{CHE05}%
  \BibitemOpen
  \bibfield  {author} {\bibinfo {author} {\bibfnamefont {J.}~\bibnamefont
  {Chen}}, \bibinfo {author} {\bibfnamefont {J.~S.}\ \bibnamefont {Thorp}}, \
  and\ \bibinfo {author} {\bibfnamefont {I.}~\bibnamefont {Dobson}},\
  }\href@noop {} {\bibfield  {journal} {\bibinfo  {journal} {Int. J. Electr.
  Power Energy Syst.}\ }\textbf {\bibinfo {volume} {27}},\ \bibinfo {pages}
  {318} (\bibinfo {year} {2005})}\BibitemShut {NoStop}%
\bibitem [{\citenamefont {Nedic}\ \emph {et~al.}(2006)\citenamefont {Nedic},
  \citenamefont {Dobson}, \citenamefont {Kirschen}, \citenamefont {Carreras},\
  and\ \citenamefont {Lynch}}]{NED06}%
  \BibitemOpen
  \bibfield  {author} {\bibinfo {author} {\bibfnamefont {D.~P.}\ \bibnamefont
  {Nedic}}, \bibinfo {author} {\bibfnamefont {I.}~\bibnamefont {Dobson}},
  \bibinfo {author} {\bibfnamefont {D.~S.}\ \bibnamefont {Kirschen}}, \bibinfo
  {author} {\bibfnamefont {B.~A.}\ \bibnamefont {Carreras}}, \ and\ \bibinfo
  {author} {\bibfnamefont {V.~E.}\ \bibnamefont {Lynch}},\ }\href@noop {}
  {\bibfield  {journal} {\bibinfo  {journal} {Int. J. Electr. Power Energy
  Syst.}\ }\textbf {\bibinfo {volume} {28}},\ \bibinfo {pages} {627} (\bibinfo
  {year} {2006})}\BibitemShut {NoStop}%
\bibitem [{\citenamefont {Dobson}\ \emph {et~al.}(2005)\citenamefont {Dobson},
  \citenamefont {Carreras},\ and\ \citenamefont {Newman}}]{DOB06}%
  \BibitemOpen
  \bibfield  {author} {\bibinfo {author} {\bibfnamefont {I.}~\bibnamefont
  {Dobson}}, \bibinfo {author} {\bibfnamefont {B.~A.}\ \bibnamefont
  {Carreras}}, \ and\ \bibinfo {author} {\bibfnamefont {D.~E.}\ \bibnamefont
  {Newman}},\ }\href@noop {} {\bibfield  {journal} {\bibinfo  {journal}
  {Probab. Eng. Inform. Sc.}\ }\textbf {\bibinfo {volume} {19}},\ \bibinfo
  {pages} {15} (\bibinfo {year} {2005})}\BibitemShut {NoStop}%
\bibitem [{\citenamefont {Dobson}\ \emph {et~al.}(2004)\citenamefont {Dobson},
  \citenamefont {Carreras},\ and\ \citenamefont {Newman}}]{DOB04}%
  \BibitemOpen
  \bibfield  {author} {\bibinfo {author} {\bibfnamefont {I.}~\bibnamefont
  {Dobson}}, \bibinfo {author} {\bibfnamefont {B.~A.}\ \bibnamefont
  {Carreras}}, \ and\ \bibinfo {author} {\bibfnamefont {D.~E.}\ \bibnamefont
  {Newman}},\ }in\ \href@noop {} {\emph {\bibinfo {booktitle} {37th Hawaii
  International Conference on System Sciences}}}\ (\bibinfo {address}
  {Hawaii},\ \bibinfo {year} {2004})\BibitemShut {NoStop}%
\end{thebibliography}%

\clearpage

\section{Appendix}

\noindent\textbf{1. Distribution of $p_c$ around the mean-field prediction.}

\begin{figure}[h]
 \includegraphics[width=0.67\columnwidth]{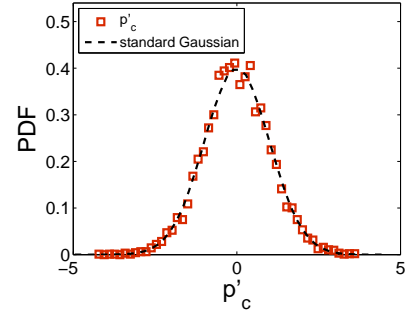}
 \caption{(Color online) PDF of the normalized $p_c$, $p'_{c}$, for single realizations in simulations, compared with the standard Gaussian distribution. Here, $k=5$, $N=30\,000$ for $6000$ realizations. We {see here} that the value of $p_c$ follows {quite well} a Gaussian distribution.}
 \label{fig7}
\end{figure}

{Fig. \ref{fig7} shows the PDF of the normalized values of $p_c$:
$p'_{c}\equiv\frac{p_c-\langle p_c\rangle}{\textup{\textbf{std}}(p_c)}$,
compared with a standard Gaussian distribution. Here we can find that $p_c$ follows a Gaussian
distribution around the MF prediction $p_{c}^{MF}$. This supports our assumption in the main
text that $p_{c}$ follows a Gaussian distribution.}

\

\noindent\textbf{2. Effect of the randomness in network structures}

\begin{figure}[h]
 \includegraphics[width=0.67\columnwidth]{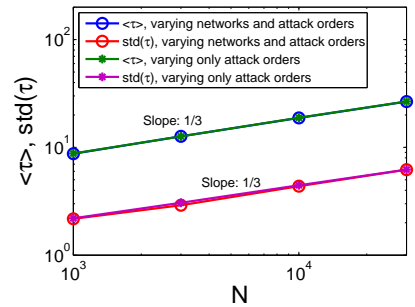}
 \caption{{(Color online) Scaling behaviors of the mean and the standard deviation of the total
 time $\tau$ at $p_{c}$ for individual realizations. Two cases
 are compared here: in each realization, varying both the networks and the attack order, and only
 varying the attack order. We consider $k=5$ with $3000$ realizations for different $N$ values.}}
 \label{fig8}
\end{figure}

{In our simulations, there are two types of randomness in each realization: the structure of
ER network and the random initial attack. We always change both the networks and the attack order
at the beginning of each realization. However, when the network is large enough, the randomness of
the network structure is not needed for our results. In Fig. \ref{fig8}, we compare the scaling
behaviors of the total number of cascade $\tau$ in two cases: varying both the networks and the attack
order, and varying only the attack order for a given realization. We find that they have very similar behavior.}

\end{document}